\documentclass[preprint,12pt,sort&compress]{elsarticle}

\usepackage{amssymb}
\usepackage{amsmath}
\usepackage{booktabs}
\usepackage{multirow}
\usepackage{xcolor}
\usepackage{url}
\journal{Ocean Engineering}

\begin{document}

\begin{frontmatter}



\title{Operator Learning for Predicting Bulk Wave Parameters of Spectral Wave Models}


\author[authorLabel1]{Shukai Cai\corref{cor1}}
\cortext[cor1]{Corresponding author}
\ead{shukai.cai@utexas.edu}

\author[authorLabel1]{Sourav Dutta}
\author[authorLabel2]{Mark Loveland}
\author[authorLabel3,authorLabel4]{Eirik Valseth}
\author[authorLabel2]{Peter Rivera-Casillas}
\author[authorLabel2]{Corey Trahan}
\author[authorLabel1]{Clint Dawson}%% Author name

\affiliation[authorLabel1]{organization={Oden Institute for Computational Engineering and Sciences, The University of Texas at Austin},%Department and Organization
            addressline={201 E 24th St}, 
            city={Austin},
            postcode={78712}, 
            state={TX},
            country={USA}}
\affiliation[authorLabel2]{organization={Information Technology Laboratory, U.S. Army Engineer Research and Development Center},%Department and Organization
            % city={},
            state={MS},
            country={USA}}
\affiliation[authorLabel3]{organization={Department of Mechanical Engineering and Technology Management, Norwegian University of Life Sciences},%Department and Organization
            % city={},
            state={Ås},
            country={Norway}}
\affiliation[authorLabel4]{organization={Department of Numerical Analysis and Scientific Computing, Simula Research Laboratory},%Department and Organization
            % city={},
            state={Oslo},
            country={Norway}}

%% Abstract
\begin{abstract}
The impact of wave-induced forcing on the mean water level and nearshore currents is typically modeled through excess momentum fluxes, also known as radiation stresses, and their spatial gradients. Accurate storm surge prediction requires coupled circulation and wave models, but the high computational cost of numerical wave models limits their temporal resolution. In this work, we explore a proof-of-concept application of Deep Operator Networks (DeepONets) as a surrogate for the Simulating WAves Nearshore (SWAN) numerical wave model. Unlike grid-dependent surrogate models, DeepONets learn the underlying continuous operator, and thus, can provide highly efficient prediction while enabling discretization-invariant inference. The proposed surrogate model is evaluated using two distinct 1-D and 2-D steady-state numerical examples with variable boundary wave conditions and wind fields. When applied to a realistic numerical example of steady-state wave simulation in Duck, NC, the DeepONet surrogate improves computational efficiency by four orders of magnitude. Furthermore, the model demonstrates consistently high accuracy in predicting the significant wave height and the x- and y- components of the radiation stress gradient, by achieving relative $L_2$ errors bounded by $1.91\%$, $10.98\%$, and $6.88\%$, respectively, across all unseen test scenarios.

\end{abstract}

% \includegraphics{GraphicalAbstract.eps}

%% Keywords
\begin{keyword}
Deep Operator Network \sep Surrogate Modeling \sep Wave-Current Interactions \sep Simulating Waves Nearshore




\end{keyword}

\end{frontmatter}

% \linenumbers

\setcounter{section}{0}
\section{Introduction}\label{sec1}
Wind-driven ocean waves are a fundamental physical process that mediates the exchange of momentum, heat, and gases between the atmosphere and the ocean, directly influencing global climate patterns and coastal morphology. Beyond their role in Earth's energy budget, these waves dictate the safety of maritime operations and the structural integrity of offshore infrastructure. Numerical modeling of wind waves using spectral models is notoriously difficult because it requires solving the Wave Action Balance Equation (WABE), which tracks the wave action density across multiple dimensions, namely the spatio-temporal evolution across vast, irregular oceans, as well as the spectral distribution of energy across different frequencies and directions~\cite{booij1999third,holthuijsen2010waves}. In particular, accurate modeling of wave-wave interactions, represented as a nonlinear source term, involves complex multidimensional integrals that are usually computationally intractable for real-time simulations or long-term projections. 

In addition, wave setup exhibits a significant nonlinear interaction with storm tides, forming a complex system of wave–current interactions. Therefore, accurate representation of these processes is essential to improve the accuracy of storm surge predictions. Causio et al.~\cite{os-21-1105-2025} showed that wave-induced surge components may contribute $10\%$ - $30\%$ to the total water level during storms. This wave-induced contribution originates from the transfer of momentum from surface waves to currents, which occurs through radiation stress~\cite{longuet1962radiation}. As a result, water levels and flow patterns can be substantially impacted, particularly during hurricane events. 
To capture these effects, researchers have coupled circulation models with wave models, such as Simulating WAve Nearshore + ADvanced CIRCulation (SWAN+ADCIRC) and Semi-implicit Cross-scale Hydroscience Integrated System Model + Wind Wave Model-III (SCHISM+WWM3)~\cite{dietrich2011modeling, hsiao2019quantifying}. 
However, due to their intrinsic complexity and computational cost, phase-averaged spectral wave models typically account for a significant portion of the overall computational cost in these coupled wave-current simulations. 

A wave model typically accounts for physical processes governing the generation, transformation, and dissipation of ocean waves. Wave growth is primarily driven by wind forcing, while energy is dissipated through mechanisms such as whitecapping, bottom friction, and depth-induced breaking~\cite{phillips1957generation, miles1957generation, hasselmann1974spectral, shemdin1978nonlinear, battjes1978energy}. In addition, nonlinear wave–wave interactions redistribute energy across the spectrum, and wave–current interactions further modify wave evolution~\cite{eldeberky1996statistical, hasselmann1974spectral}. These processes are strongly influenced by environmental factors, including local bathymetry, coastline geometry, and spatially varying wind conditions. As a result, wave evolution is governed by strong nonlinear couplings across physical and spectral scales. This complexity significantly increases the computational cost of numerical wave simulation, particularly in coupled wave-current modeling frameworks. At the mathematical level, this complexity is reflected in the form of the governing equations used in spectral wave models. 

The governing equations of spectral wave models, i.e., the WABE, are high-dimensional partial differential equations that describe the evolution of wave action density in both physical and spectral space. Typically, these equations are numerically approximated using finite difference, finite element, and finite volume methods to simulate wave propagation over large spatial domains. The development of one of the earliest such models, the WAve Model (WAM), began in the early 1970s and evolved through three generations, as summarized by the WAMDI Group~\cite{group1988wam}. First-generation models prescribed spectral shapes with empirical growth laws, whereas second-generation models partially evolved the spectrum but required imposed constraints on nonlinear interactions. In third-generation models, the wave action density is computed at each computational grid point, greatly enhancing accuracy and physical consistency. However, this increased fidelity also leads to higher computational cost, which remains a challenge for large-scale problems. There are many historical review articles~\cite{janssen_2008,roland_ardhuin_2014} that summarize the methods employed in contemporary numerical spectral wind wave models, including some of the most well-known models used in practice such as ECWAM~\cite{janssen2004_ecwam}, SWAN~\cite{booij_ris_holthuijsen_1999_swan}, and WAVEWATCH III~\cite{tolman_1991_ww3}. 

The persistent computational burden in these third-generation wave models has motivated growing interest in alternative, data-driven modeling strategies that can retain physical fidelity while enabling rapid prediction. These methods leverage the increasing availability of data and advances in machine learning to enhance traditional modeling techniques. In \cite{jorges2023spatial}, the authors argued that most studies applying machine learning to ocean wave height prediction focus on single-point predictions. For problems in two-dimensional (2-D) spatial domains, convolutional neural networks (CNNs) are popular due to their ability to learn relevant spatial correlations in image-type data on regular grids~\cite{lecun2015deep}. CNNs have been shown to be successful in predicting the significant wave height (Hsig) for both regional ocean and experimental study domains, with rectangular geometries and regular computational meshes~\cite{bai2022development, choi2020real, wei2022convolutional}. For a specific coastal domain of the North Sea using SWAN-generated wave fields as ground truth, CNN-based surrogate models achieved an average spatial root mean square error (RMSE) of 0.097 m for Hsig \cite{jorges2023spatial}. In addition to CNNs, Feed-Forward Neural Networks and Long Short-Term Memory (LSTM) networks have also been applied to predict the time series of Hsig evolution for station-based locations~\cite{feng2020multi, fan2020novel}.

Despite these successes, existing data-driven approaches face several fundamental limitations when applied to realistic coastal wave modeling problems. CNN-based models are not directly applicable to complex geometries or unstructured meshes. Moreover, their inference capability is inherently restricted to the same spatial discretization and grid configuration as the training data. LSTM-based models typically also suffer from this limitation as they often require a spatial embedding of the data for computational efficiency or are restricted to point-based predictions.

In contrast, Deep Operator Networks (DeepONets) belong to a specialized class of deep neural network-based architectures known as neural operators (NO), that are designed to approximate mappings between infinite-dimensional functional spaces~\cite{lu2019deeponet}, and are inherently independent of the spatial discretization of the output function. These properties enable neural operator frameworks such as DeepONets to efficiently approximate the solution operator of complex PDEs. Hence, DeepONet-based architectures have been tremendously successful in a wide array of computational science and engineering applications~\cite{osorio_wang_2022,mao_dong_perdikaris_2023,shashank_kushwaha_2024,diab_kobaisi_2024,mitonet_2026}. In the context of ocean and wave modeling, the application of DeepONet to model high-raft wave energy converter (WEC) by Zhang et al.~\cite{zhang_zhao_2023} was the first study to demonstrate the feasibility and superiority of operator learning in predicting the response of floating structures.  Subsequently, Cao et al.~\cite{cao2024deep} investigated the capability of three neural operator architectures — DeepONets, Fourier Neural Operator (FNO), and Wavelet Neural Operator (WNO) to model floating offshore structures under irregular ocean waves, and also found that DeepONets with historical states performed best for broadband and transient responses. Zhang et al.~\cite{zhang2025multiple} applied a variant of DeepONet called the Multiple-Input Operator Network (MIONet) to successfully simulate the nonlinear dynamic response of a bistable WEC driven by an irregular wave elevation time series. Zheng et al.~\cite{zheng2025phase} used multi-scale FNO for phase-resolved wave prediction from radar observations and achieved a $14\%$ reduction in prediction mean absolute error compared to the standard FNO. Zhang et al.~\cite{zhang2025deep} utilized the DeepONet framework to predict future wave heights using historical wave height data. In a controlled laboratory wave basin setup, their model showed impressive accuracy in short-term prediction horizons over a range of sea states (WC01-WC04), and outperformed an LSTM-based wave prediction model by about $30\%$ across both short- and medium-term prediction horizons. 

This study aims to explore and demonstrate the applicability of the DeepONet framework as a data-driven surrogate for spectral wind wave models like SWAN, capable of approximating the relationships between wind forcing and boundary wave conditions with the steady-state conditions of the sea surface in realistic coastal domains. In particular, this study investigates the feasibility of using DeepONets to predict Hsig and radiation stress gradients, two key variables used to characterize the sea state in spectral wave models and quantify wave-current interactions in numerical models of the tightly coupled wave-current system, such as the SWAN+ADCIRC model~\cite{dietrich2011modeling}. Thus, this study aims to establish a foundation for the development of NO-based accelerated frameworks for coupled wave-current models as a potential future direction of research. 

Despite the similarity in model architecture with ~\cite{zhang2025deep}, there are several key differences in the conceptual design and functionality. Unlike in ~\cite{zhang2025deep}, where a DeepONet is used to learn the time series of wave height evolution using historical wave height observations, this study focuses on learning the relationship between the steady-state wave state and the driving factors behind it such as boundary conditions and body forces. First, this enables the proposed modeling framework to accurately represent a wide range of ambient environmental conditions. Second, this modeling framework more closely emulates the solution operator underlying spectral wind wave models such as SWAN, particularly in steady-state mode, and would, therefore, admit a natural extension to a surrogate wave model that can be deployed in a coupled wave-current system in the future. Moreover, the proposed framework preserves DeepONet's property of discretization invariance in the output function space, and thus could be used to predict Hsig and the gradient of radiation stresses even at locations not included in the computational mesh of the numerical wave model-generated training data.

The remainder of the paper is organized as follows: Section~\ref{sec2} introduces the physical background and numerical framework, including the SWAN wave model, wave-current coupling, and the DeepONet architecture used to learn nonlinear mappings from wind and wave conditions to spatially distributed wave quantities. Section~\ref{performance_indicators} defines the multi-scale performance indicators used to evaluate the surrogate, from global scenario-based metrics to node-wise spatial error distributions. Section~\ref{sec3} presents numerical results for two examples: a one-dimensional (1-D) domain with uniform-slope bathymetry, a subsequent sampling size sensitivity analysis evaluating computational training cost, and a 2-D realistic coastal domain (DUCK), based on experiments conducted at the Field Research Facility (FRF) of the U.S. Army Engineer Research Development Center's Coastal and Hydraulics Laboratory, located in Duck, North Carolina~\cite{Delilah_1997}, and which has been included in the SWAN testbed~\cite{ris2003onr}. The section also includes cross-grid verification to evaluate discretization invariance. Section~\ref{sec4} discusses the physical interpretability of the surrogate model, and Section~\ref{sec5} summarizes the main findings and outlines directions for future work.

\section{Methods}
\label{sec2}
\subsection{Problem Definition}

The physical quantities of interest in this study are (a) the Hsig, which represents a phase-averaged statistical measure for estimating the evolution and dissipation of the wave energy, and (b) the radiation stress gradients, which represent the wave-induced momentum flux transferred from surface waves to the current. Using Hsig as a modeled output for a spectral wave surrogate model is highly effective because it provides a low-dimensional, high-signal representation of complex spectral data, making it easier for machine learning architectures to converge on physically consistent results. Additionally, since Hsig is the industry standard for coastal risk assessment and defines non-linear thresholds like wave breaking, modeling it directly ensures that the surrogate provides immediate, actionable insights for engineering and validation. On the other hand, in coupled wave-current models such as SWAN+\allowbreak ADCIRC, radiation stress gradients appear in the current model (ADCIRC) as an additional forcing term. These gradients are computed by SWAN from the wave field and supplied to ADCIRC.

SWAN solves the wave action balance equation (WABE) defined as:
\begin{equation}
    \text \quad \frac{\partial N}{\partial t} + \nabla_{\vec{x}} \cdot \left[(\vec{c}_g + \vec{U}) N \right] + \frac{\partial c_{\theta} N}{\partial \theta} + \frac{\partial c_{\sigma} N}{\partial \sigma} = \frac{S}{\sigma}.
\end{equation}
Here, \( N(\sigma, \theta, x, y, t) \) represents the action density spectrum, which quantifies the wave energy per unit frequency and direction. It is defined as \( N = \frac{E}{\sigma} \), where \( E \) denotes the energy density of the waves, and \( \sigma \) is the relative frequency. On the left-hand side, the first term captures the rate of change of the action density over time. The second term is the spatial propagation term, which represents the propagation of the action density in geographic space \(\vec{x}\) with the wave group velocity \(c_{g}\) and the ambient current vector \(\vec{U}\). The third term is the spectral propagation term, which represents depth-induced / current-induced refraction with propagation velocity \(c_{\theta}\) in terms of propagation direction \(\theta\), and the fourth term is the change in the relative frequency due to variations in depth and currents with propagation velocity \(c_{\sigma}\) in terms of relative frequency \(\sigma\). The source term, $S$ on the right-hand side represents the combined effects of wind energy input, nonlinear wave-wave interactions, and various dissipation processes. SWAN incorporates these source terms using empirical formulations as documented in the SWAN Scientific and Technical Documentation~\cite{booij1999third,swan_manual}. 

To estimate the characteristic height of random waves in a sea state, a standardized statistic called the Hsig is computed from the wave energy density spectrum as
\begin{equation}
{\text{Hsig}} = 4\sqrt{\int_0^\infty \int_0^{2\pi} E(\sigma,\theta) \,d\theta \, d\sigma}.
\end{equation}
This is an important bulk wave parameter estimated by spectral wave models as it corresponds to the average wave height visually estimated by a mariner, and is typically also reported by weather forecast and buoy measurements. 

To simulate the complex interactions between waves and circulation in coastal environments, SWAN is often coupled with ADCIRC, a state-of-the-art high-performance numerical model for free-surface circulation, tides, and storm surge.  ADCIRC~\cite{westerink_1992} solves a form of the 2-D shallow water equations, where the continuity equation is reformulated as a generalized wave continuity equation~\cite{lynch1979wave}. The resulting system of equations is solved using a Bubnov-Galerkin finite element method~\cite{dawson2006continuous} with linear polynomial basis functions. Within the generalized wave continuity and momentum equations, radiation stress gradients appear explicitly as additional forcing terms that modify the momentum balance, thereby representing the net effect of wave-induced forcing on the current field. 
These wave-induced forcing components per unit surface area ($F_x$ and $F_y$) are defined through the spatial gradients of the radiation stress tensor components:
\begin{equation}\label{eq:forces_definition}
F_x = - \frac{\partial S_{xx}}{\partial x} - \frac{\partial S_{xy}}{\partial y}, \quad F_y = - \frac{\partial S_{yx}}{\partial x} - \frac{\partial S_{yy}}{\partial y},
\end{equation}
Throughout this study, the terms “wave-induced forces” and “radiation stress gradients” are used interchangeably to denote $F_x$ and $F_y$, the x- and y-force components, respectively. where the underlying components of the radiation stress tensor ($S_{xx}, S_{yy}, S_{xy}$) are integrated across the directional wave energy spectrum $E(\sigma, \theta)$ according to linear wave theory:
\begin{align}
S_{xx} &= \rho g \int_0^\infty \int_0^{2\pi} \left( n - \frac{1}{2} + n \cos^2 \theta \right) E(\sigma, \theta) \, d\theta \, d\sigma,\\
% \end{equation}
%
% \begin{equation}
S_{yy} &= \rho g\int_0^\infty \int_0^{2\pi} \left( n - \frac{1}{2} + n \sin^2 \theta \right) E(\sigma, \theta) \, d\theta \, d\sigma,\\
% \end{equation}
%
% \begin{equation}
S_{xy} &= \rho g\int_0^\infty \int_0^{2\pi} n \sin \theta \cos \theta \, E(\sigma, \theta) \, d\theta \, d\sigma.
\end{align}
Here, $\rho$ is the water density, $g$ is the constant of gravitational acceleration, and $n$ represents the fraction of wave group velocity to phase velocity.

Therefore, at each coupling step, SWAN updates the wave spectrum pointwise across the spatial computational grid and derives the associated radiation stress. Because this process involves solving a high-dimensional spectral problem under dynamically updated circulation conditions across a large geographic domain, the computational cost of running SWAN at every coupling time step is expensive. 
Hence, this study investigates the feasibility of using DeepONet not only as a surrogate model for SWAN to estimate bulk parameters such as Hsig, but also to improve the efficiency of predicting radiation stress gradients without compromising accuracy across varying conditions. 

\subsection{DeepONet Architecture}
\label{DeepONetArchitecture}

DeepONet is an operator learning framework proposed by Lu et al.~\cite{lu2019deeponet}, based on the universal approximation theorem for operators~\cite{chen1995universal}. Unlike traditional deep neural networks that approximate mappings between finite-dimensional tensors, DeepONets are designed to approximate a nonlinear mapping between two infinite-dimensional function spaces, which we denote as: $\mathcal{\hat{G}}: \mathcal{U} \mapsto \mathcal{V}$, where $\mathcal{U}$ represents the space of input functions and $\mathcal{V}$ denotes the space of output functions.

In the context of physics-based modeling, the target operator $\mathcal{G}$ is implicitly defined by the governing equations of the underlying system. Operator learning aims at building an approximation $\hat{\mathcal{G}}(\boldsymbol{\theta}) : \mathcal{U} \mapsto \mathcal{V}$, $\boldsymbol{\theta}$ being the parameters of the neural operator, such that $\hat{\mathcal{G}}(u, \boldsymbol{\theta}) \approx \mathcal{G}(u)$ for any input function $u \in \mathcal{U}$. In practical applications, the training data consist of input--output function pairs $(u(\mathbf{x}), v(\mathbf{y}))$, where $u \in \mathcal{U}$ and $v \in \mathcal{V}$, evaluated at discrete coordinates $\mathbf{x} \in \mathcal{X}$ and $\mathbf{y} \in \mathcal{Y}$. Consequently, the operator mapping can be interpreted pointwise such that $v(\mathbf{y}) = \mathcal{\hat{G}}(u)(\mathbf{x})$.

\begin{figure}[h!]  
    \centering
    \includegraphics[width=0.9\textwidth]{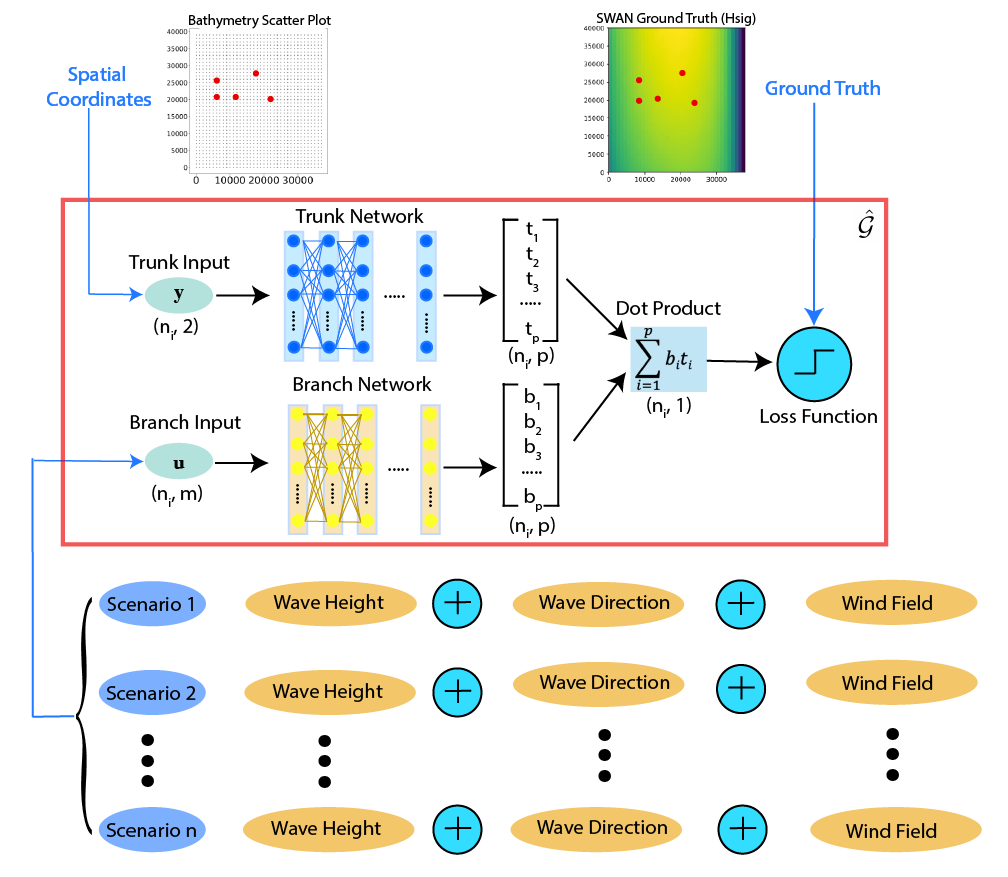}  
    \caption{Conceptual overview of the DeepONet architecture for modeling bulk wave parameters. For each wave–wind scenario, boundary wave information and wind forcing are encoded by the branch network, while spatial coordinates for evaluating the output function are encoded by the trunk network to their corresponding $p$-dimensional vector fields. The DeepONet approximation of the solution function at the queried coordinates is obtained by taking a dot product of the branch and trunk outputs, thus implicitly approximating the SWAN solution operator. The high-fidelity SWAN solutions are used as the ground truth to supervise training through a point-wise loss function.}
    \label{fig:don_architecture}  % Label for referencing
\end{figure}

To efficiently approximate this operator, the DeepONet architecture utilizes two sub-networks: a branch network $\boldsymbol{f}$ and a trunk network $\boldsymbol{g}$, each responsible for encoding a different aspect of the operator, as illustrated in Figure~\ref{fig:don_architecture}.

The branch network encodes the input function $u$, which can correspond to boundary conditions and other forcing inputs, through its evaluations at a set of representative sampling points $\{\mathbf{x}_i \in \mathcal{X}\}_{i=1}^m$, given by $\{u_1 = u(\mathbf{x}_1), u_2= u(\mathbf{x}_2), \ldots, u_m = u(\mathbf{x}_m)\}$. The output of the branch network is a feature vector $\boldsymbol{b} \in \mathbb{R}^p$.
% , which represents the global dependence of the solution on the input function. 
%The branch network architecture is flexible and can be chosen based on the application.
% 
The trunk network, on the other hand, encodes the coordinates $\mathbf{y} \in \mathcal{Y}$ at which the output function is evaluated. Its output is a feature vector $\boldsymbol{t} \in \mathbb{R}^p$, with the same dimensionality as the output of the branch network. The DeepONet framework allows flexibility in choosing neural network architectures for the branch and trunk networks based on the application's requirements.

The branch and trunk network outputs are combined through an inner product to produce an approximation of the output function value at the specified coordinate, $\mathbf{y}$. The loss function is defined as the mean squared error between the predicted and ground-truth values at the queried coordinates. By minimizing the loss function during training, the surrogate model learns an accurate continuous approximation of the underlying physical operator. Once trained, the output for an unseen input function can be obtained at any arbitrary coordinate $\mathbf{y}^*$ by providing the corresponding discretized input function values at the fixed representative sampling points $\mathbf{u}^* = \{u^*_1, u^*_2, \ldots, u^*_m \}$ as the branch input and the queried coordinate $\mathbf{y}^*$ as the trunk input, to produce the model prediction as $\hat{\mathcal{G}}(\mathbf{u}^*)(\mathbf{x}^*)$.
Consequently, even though the sampling points at which the input function is evaluated remain fixed during training and inference, the DeepONet framework allows evaluation of the learned operator at arbitrary coordinates, independent of the training mesh. In this work, we adapt the DeepONet architecture to study the abstract nonlinear functional mapping from boundary and field conditions to bulk wave parameters, such as the Hsig and the radiation stress gradients. 

\subsection{Operator Formulation}

In this study, SWAN is used as a high-fidelity, physics-based solver to generate training and testing data from simulations of two distinct numerical examples. To focus on the feasibility of learning the nonlinear functional map from wind and boundary wave conditions to the fully developed wave state, steady-state SWAN simulations are utilized for all numerical examples in this study. This is a common practice in coastal engineering studies to reduce computational cost and is appropriate for wave conditions that vary more slowly than the time it takes for waves to propagate through the domain or to attain fetch-limited or fully developed conditions~\cite{STWAVE_2011}. A Joint North Sea Wave Project (JONSWAP) spectrum~\cite{hasselmann1973_jonswap} is prescribed at the boundary, and wave height and peak wave direction are considered input variables. The wind field (wind direction and velocity) across the domain is also considered an input to the surrogate model.

From a systems perspective, the resulting target operator that the DeepONet model learns in this study can be formulated as an implicit nonlinear map between the space of boundary wave conditions and wind forcing functions to the space of the bulk wave parameters, such as $H_{sig}, F_x,$  and $F_y$, and written formally as:
\begin{equation}
\mathcal{G}_S:\ \bigl(W_{\mathrm{bnd}},\ \mathbf{U}_{10})\ \longmapsto\ \{H_{sig}, F_x, F_y\}(\mathbf{y}),
\label{eq:swan_operator}
\end{equation}
where $\mathcal{G}_S$ denotes the operator map, $W_{\mathrm{bnd}}$ denotes the prescribed boundary wave conditions (wave height, direction, etc.), $\mathbf{U}_{10}$ represents the wind field, and $\mathbf{y}$  represents discrete coordinates at which the output field is queried. It is important to note that this operator map formulation can, in principle, be extended to any other combination of input features and target functions of interest to the wave modeling community, provided that there is a physics-guided or causal relationship between the mapped quantities.

\subsection{Numerical Examples}\label{sec:scenario_setup}

This study investigates the application of DeepONet for wave simulation using two distinct numerical examples. The first example involves simplified configurations with a one-dimensional (1-D) domain, where the bathymetry is modeled as a uniform slope. The simplified example serves as a controlled environment for examining DeepONet's ability to learn fundamental aspects of wave propagation. The second example is derived from the F71 test case in the ONR Test Bed~\cite{ris2003onr}, which is intended as a benchmark for wave models in nearshore field experiments with depth-induced breaking. The F71 test case also includes real-world pressure gage data from the DELILAH nearshore experiment (Duck Experiment on Low-frequency and Incident-band Longshore and Acrossshore Hydrodynamics), which was conducted at the Field Research Facility (RFR) on the shores of the Atlantic Ocean in Duck, North Carolina, USA. The steady-state numerical example derived from this test case, henceforth referred to as the DUCK example, provides validation for applying DeepONets to realistic coastal environments. Unlike the simplified 1-D uniform-slope bathymetries, the DUCK example involves the real-world bathymetry of the FRF area, complex boundary conditions, and realistic coastal wind conditions. Together, these two numerical examples provide an extensive evaluation framework, ranging from idealized configurations to a realistic coastal environment, to assess the capability and generalization properties of the proposed DeepONet surrogate. 

For all the examples, SWAN simulations are first performed to generate the training and testing datasets. To ensure data quality, the following convergence criteria are applied to each simulation. SWAN terminates the iterations when the relative change and curvature conditions of Hsig are satisfied, and these criteria hold true for more than $99.5\%$ of the wet grid points. If a simulation does not reach the convergence criterion within $50$ iterations, it is excluded from the data set.

\begin{figure}[h!]
\centering
\includegraphics[width=0.9\textwidth]{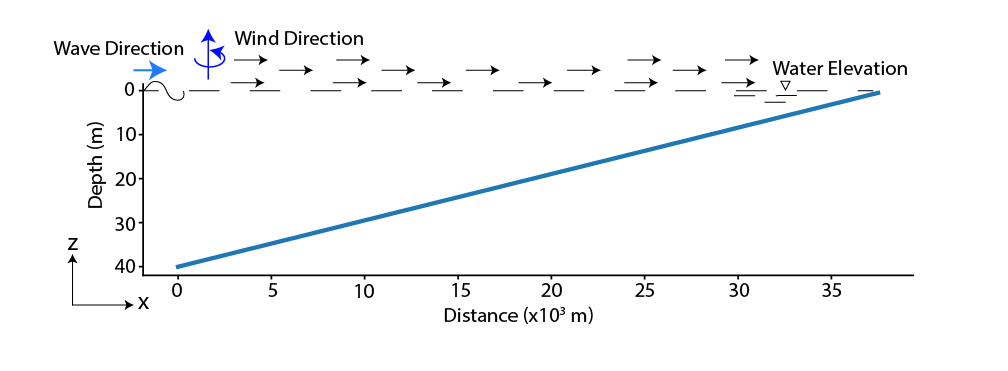}
\caption{Schematic diagram of the 1-D example. The wave condition at $x=0$ and propagates along the positive x-axis.}
\label{fig:1d_case_depth}
\end{figure}

As a first step, we evaluate the performance of the proposed DeepONet surrogate using an idealized 1-D example with uniform-slope bathymetry, which provides a controlled setting for assessing basic wave propagation and transformation behavior. As shown in Figure \ref{fig:1d_case_depth}, the depth decreases linearly from 40 m to 0 m in the x-direction over a distance of 38 km, resulting in a bathymetry gradient of approximately 0.11, corresponding to an inclination angle of approximately $6.3^\circ$. Based on this simplified geometric configuration, $2401$ parametric scenarios are simulated using SWAN by varying the wind speed, wind direction, wave height, and wave direction as detailed in Table~\ref{tab:Variables_for_SWAN}. The JONSWAP spectrum is applied at the offshore boundary at $x=0$, and the wave propagates along the positive x-direction. The JONSWAP spectrum is defined with a peak period of $3.5$ s and a directional spreading power of $2.0$. This parametric dataset is split into 1,680 training, 360 validation, and 361 test scenarios (a 70-15-15 split). To ensure proper distribution and prevent bias in the dataset, the scenarios are shuffled before splitting.

\begin{table}[ht]
\centering
\resizebox{\textwidth}{!}{
\begin{tabular}{c c c c}
    \toprule
    &  \textbf{Parameter} & \textbf{1-D Example} & \textbf{DUCK Example} \\
    \midrule
    & Reference Frame & Positive x-axis = $0^{\circ}$ & True North (+y) = $0^{\circ}$\\
    & Direction Convention & Propagation Direction (toward) & Nautical Direction (from)\\
    \midrule
    \multirow{4}{*}{\rotatebox{90}{Wind}} & Wind Speed  & $11-17~m/s$ & $1-8~m/s$ \\
    & Increment  &$1~m/s$ &$1~m/s$ \\
    & Wind Direction  &$-7^\circ$ to $7^\circ$ &$0^\circ$ to $315^\circ$ \\
    & Increment  &$1^\circ$ &$45^\circ$ \\
    \midrule
    \multirow{4}{*}{\rotatebox{90}{Wave}} & Boundary Hsig  & $0.6-1.2~m$ & $1.6-2.6~m$ \\
    & Increment  &$0.1~m$ &$0.2~m$ \\
    & Boundary Wave Direction  &$-7^\circ$ to $7^\circ$ &$-20^\circ$ to $160^\circ$ \\
    & Increment  &$1^\circ$ &$20^\circ$ \\
    \bottomrule
\end{tabular} 
}
\caption{Summary of input parameter spaces, coordinate reference frames, and directional conventions utilized for generating the training and testing datasets in the 1-D uniform slope and 2-D DUCK field experiment examples.}\label{tab:Variables_for_SWAN}\label{tab:Variables_for_SWAN}%
\end{table}

\begin{figure}[ht]  
\centering
\includegraphics[width=0.9\textwidth]{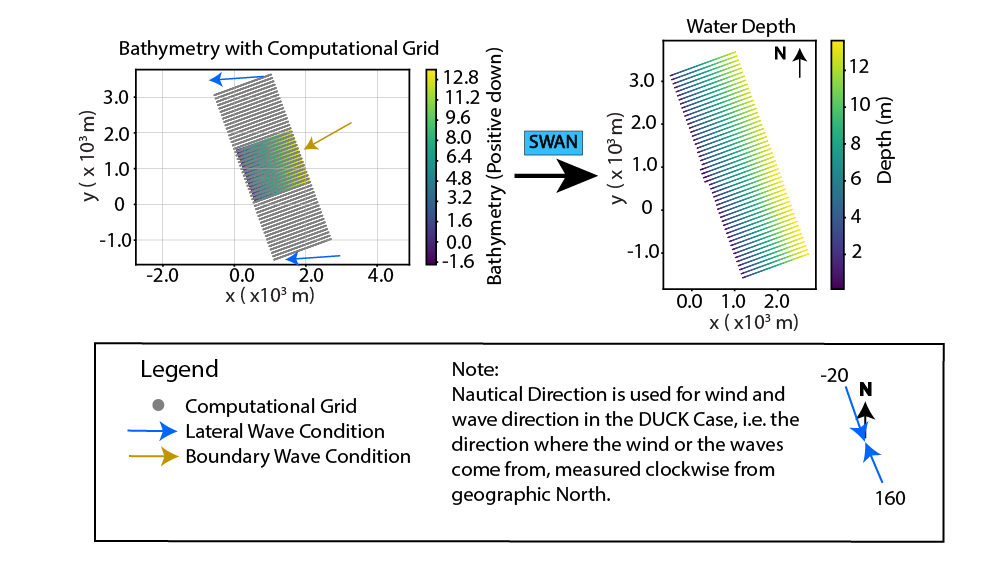}  
\caption{The left figure shows the overlap of the computational grid centers (gray) with the bathymetry (color). The computational grid extends beyond the surveyed bathymetry to minimize boundary effects. After SWAN is run, it generates the x, y coordinates along with their corresponding water depths, Hsig, and other target wave properties. The figure on the right illustrates the spatial distribution of water depth.}
\label{fig:duck_depth} 
\end{figure}

In the DUCK example, the bathymetry and computational grid are obtained from the F71 test case of the ONR Test Bed~\cite{ris2003onr}, as illustrated in Figure \ref{fig:duck_depth}. The computational domain is defined as a rectangular region, spanning $1700$ m by $5000$ m in the cross-shore and along-shore directions, respectively, with the origin at ($1128.66$m, $-1597.47$m), and rotated counterclockwise by $20$ degrees relative to the true East. This is discretized into a $136 \times 50$ regular structured grid, corresponding to a uniform grid resolution of $12.5$ m in the cross-shore direction and $100$ m in the along-shore direction. The directional domains is discretized into $36$ uniform directional bins ($10$ degree resolution), while the frequency domain is discretized into $53$ frequency bins from $0.04443$ Hz to $0.6$ Hz using a logarithmic progression. The bathymetric data is obtained via interpolation from a regional depth file with a spatial resolution of $13.4532$ m. 

A constant wind field is applied across the computational domain, and boundary wave conditions are imposed on the northeast boundary. In the original F71 test case, this condition is based on a spectrum derived from field observations, whereas in our example, we use a JONSWAP spectrum with a peak period of $10.718$ s and a directional spreading power of $2.0$. To verify the impact of this modification on SWAN results, the offshore boundary condition for F71 subcase 12 was reconstructed using a JONSWAP spectrum, and the SWAN simulation was repeated. The simulated Hsig values were compared with the observations of subcase 12 at all cross-shore stations. As shown in Table~\ref{tab:hsig_comparison}, the relative differences were consistently small ($< 1\%$ on average) across all stations, with a maximum relative difference of $\approx 3.0\%$ ($0.054$m absolute difference). Within this spectrum, we vary only the boundary wave's Hsig and direction. In addition, two fixed lateral wave conditions are prescribed on the north-west and south-east boundaries, consistent with the original F71 test case setup. The lateral boundary conditions are prescribed with an Hsig of $1.63$ m, a peak period of $10.5$ s, a mean direction of $88^\circ$, and a directional standard deviation of $22^\circ$. Finally, the water level is uniformly increased by $0.11$ m throughout the domain. 

\begin{table}[ht]
    \centering
    \resizebox{\textwidth}{!}{
    \begin{tabular}{c c c c}
    \toprule
      \textbf{Cross-shore Stations (Offshore $\rightarrow$ Shore)} & \textbf{Hsig Testbed (m)} & \textbf{Hsig JONSWAP (m)} & \textbf{Relative Difference(\%)}\\
      \midrule
        SAMSON & 1.64630 & 1.64006 & 0.38 \\
        Nearshore $\#1$ & 1.80501 & 1.75083 & 3.00 \\
        Nearshore $\#2$ & 1.79415 & 1.75727 & 2.06 \\
        Nearshore $\#3$ & 1.71197 & 1.70149 & 0.61 \\
        Nearshore $\#4$ & 1.62195 & 1.62701 &  0.31 \\
        Nearshore $\#5$ & 1.42923 & 1.43856 &  0.65 \\
        Nearshore $\#6$ & 1.14880 & 1.14639 & 0.21 \\
        Nearshore $\#7$ & 0.84328 & 0.84204 & 0.15 \\
        Nearshore $\#8$ & 0.59780 & 0.60117 &  0.56 \\
        Nearshore $\#9$ & 0.40141 & 0.40714 &  1.43 \\
        \bottomrule
    \end{tabular}
    }
    \caption{Comparison of significant wave height ($H_{\mathrm{sig}}$) of the F71 test case
    obtained from the SWAN Test Bed observation-based simulation and the 
    parametric JONSWAP-based simulation, at the SAMSON station and 9 different cross-shore stations, listed from farthest to nearest to shore.}
    \label{tab:hsig_comparison}
\end{table}

A total of $3,840$ scenarios are generated using SWAN, of which $2,688$ are randomly allocated to training, $576$ to validation, and $576$ to testing (70-15-15 split). Compared to the idealized 1-D example, this configuration incorporates realistic bathymetry and complex boundary interactions, providing a physically relevant test for the proposed surrogate.

\subsection{Model Architecture Setup}

Each simulation scenario is parameterized using four physical variables that represent the ambient wind and wave conditions: wind speed and wind direction, which act as domain-wide forcing, and wave height and wave direction, which prescribe the incoming wave conditions at the boundary. These are transformed into suitable input features for the branch network. Wind speed and direction are combined to be expressed as the x- and y-components of wind velocity, and each component is scaled between $-1$ and $1$ using the maximum and minimum values of the training set. The wave height is scaled between $0$ and $1$ based on its maximum and minimum values in the training set. Finally, the mean wave direction, $\theta_{bnd}$, is expressed using its sine and cosine components. Consequently, the following five input features, namely the x-and y-direction wind components ($U_x$ and $U_y$, respectively), the wave height at the boundary ($H_{bnd}$), and the components of the mean wave direction ($\cos(\theta_{bnd})$, $\sin(\theta_{bnd})$) are combined to generate the input vector of the branch network as:
\begin{equation}
\mathbf{u}
=
\bigl[
U_x,\;
U_y,\; 
H_{bnd},\;
\cos(\theta_{\mathrm{bnd}}),\;
\sin(\theta_{\mathrm{bnd}})
\bigr].
\label{eq:branch_features}
\end{equation}
%where $U_x$ and $U_y$ denote the wind velocity components, and $H_{bnd}$ and $\theta_{\mathrm{bnd}}$ represent the boundary significant wave height and mean wave direction, respectively. 
The branch network therefore encodes scenario-level forcing information.

In contrast, the trunk network samples spatial coordinates $(x,y)$ in the computational domain where the quantities of interest (Hsig and forces) are being sought. These coordinates are independently scaled to the range $[0,1]$ in both dimensions. To generate the training data, about $5\%$ of the total number of spatial coordinates is randomly sampled from each simulation scenario, and the model is trained to approximate the quantities of interest evaluated at these sampled locations. 

Within this framework, the DeepONet is trained to learn a nonlinear operator that maps forcing conditions to spatially resolved wave responses. Specifically, the DeepONet approximates the operator:
\begin{equation}
\hat{\mathcal{G}}:\ (\mathbf{u},\mathbf{x}) \ \longmapsto\ q(\mathbf{x}),
\label{eq:deeponet_operator}
\end{equation}
where $\mathbf{x}=(x,y)$ denotes the spatial coordinates provided to the trunk network, and $q(\mathbf{x})$ represents the target wave quantities of interest, including Hsig and radiation stress–related forces. 
The network output is given by:
\begin{equation}
\hat{q}(\mathbf{x})
=
\sum_{i=1}^{p}
b_i(\mathbf{u})\,
t_i(\mathbf{x}),
\label{eq:deeponet_representation}
\end{equation}
where $[b_1(\cdot), b_2(\cdot), \ldots, b_p(\cdot)]^T \in \mathbb{R}^p$ and $[t_1(\cdot), t_2(\cdot), \ldots, t_p(\cdot)]^T \in \mathbb{R}^p$ denote the $p$-dimensional outputs of the branch and trunk networks, respectively. Here $p$ is a model hyperparameter that is optimized during training.

Finally, the model is trained by minimizing the mean-squared error between the predicted outputs $\hat{q}(\mathbf{x})$ and the corresponding ground-truth values at the $N_s$ randomly sampled spatial locations $\{\mathbf{x}_i\}_{i=1}^{N_s}$, as given by,
\begin{equation}
\mathcal{L}
=
\frac{1}{N_s}
\sum_{i=1}^{N_s}
\left\|
\hat{q}(\mathbf{x}_i)-q(\mathbf{x}_i)
\right\|_2^2.
\end{equation} 
Minimizing this loss ensures that the DeepONet learns the functional mapping between boundary wave conditions, wind forcing, and the resulting wave quantities governed by the underlying physical processes.

\begin{table}[ht]
\centering
\resizebox{\textwidth}{!}{
\begin{tabular}{l l c c}
    \toprule
    & \textbf{Hyperparameter}  & \textbf{Wave Prediction Model} & \textbf{Forces Prediction Model} \\
    \midrule
    \multirow{3}{*}{\rotatebox{90}{General}} & Initial Learning Rate  & $10^{-3}$ & $10^{-4}$\\
    & Batch Size  & 256 & 64$^{*}$\\
    & Branch/Trunk Output Shape$^{1}$ & 20 & 30\\
    \midrule
    \multirow{5}{*}{\rotatebox{90}{Branch}} & Number of Layers   & 4 & 4\\
    & Neurons Per Layer & 16 & 128\\
    & Activation Function & ELU & LeakyReLU\\
    & Regularizer & none & none\\
    & Initializer & he normal & glorot uniform \\
    \midrule
    \multirow{5}{*}{\rotatebox{90}{Trunk}}
    & Number of Layers  & 5 & 5\\
    & Neurons Per Layer & 96 & 96\\
    & Activation Function & ReLU & tanh\\
    & Regularizer & none & none\\
    & Initializer & glorot uniform & he normal\\
    \bottomrule
\end{tabular} 
}
\caption{The best-performing set of hyperparameters obtained from Optuna.\\
$^{*}$ A larger batch size (256) was used for the DUCK case to reduce computational overhead, as smaller batch sizes resulted in excessive training cost.\\
$^{1}$ This is the dimension denoted by $p$ in Section~\ref{DeepONetArchitecture}.}
\label{tab:best_hyperparams}%
\end{table}

To ensure an effective model configuration while controlling computational cost, Optuna~\cite{akiba2019optuna}, an open-source hyperparameter optimization framework, is adopted to identify optimal hyperparameter configurations for DeepONet models approximating the wave height and the gradient of the radiation force for the first 1-D numerical example (see Section \ref{sec:scenario_setup}). These optimal configurations are summarized in Table~\ref{tab:best_hyperparams}, and further details of the Optuna search space are provided in the supplementary information. Due to the similarity in the model architecture, the hyperparameter optimization process is carried out only once for each model of the 1-D numerical example, and the resulting set of ``optimal'' hyperparameters is then also used to train the corresponding models for DUCK examples. Repeating the hyperparameter optimization study for DUCK example would have incurred substantial computational costs and was therefore avoided, as preliminary experiments indicated only marginal performance gains. 

Using the training configurations presented in Table~\ref{tab:best_hyperparams}, DeepONet models are trained for $10,000$ epochs using the Adam optimizer. To dynamically manage convergence, a learning rate decay scheduling algorithm (ReduceLROnPlateau) is implemented that decreases the learning rate by $10\%$ whenever the validation loss does not improve for $100$ consecutive epochs.

\section{Performance Indicators} 
\label{performance_indicators}
The performance of the proposed DeepONet surrogate model framework is evaluated using a range of error metrics. These include scenario-based metrics, which characterize the model's performance across a set of unseen test scenarios through spatially-aggregated assessments, and node-wise metrics, which measure the spatial error distribution for each test scenario to analyze the local fidelity of model predictions. The numerical solutions generated by SWAN are considered the “ground truth” for model training and testing. Thus, the proposed DeepONet surrogate model is trained to emulate the SWAN numerical solver rather than the continuous physical solution of the WABE. 

\subsection{Scenario-based Metrics}
To provide a scale-independent assessment for each test scenario, the relative $L_2$ error (RLE) is adopted to measure the normalized discrepancy between the predicted and ground-truth fields:
\begin{equation}
RLE^{(s)}= \frac{\| \hat{H}^{(s)} - H^{(s)} \|_2}{\| H^{(s)} \|_2} = \frac{\sqrt{\sum_{k=1}^{N} (\hat{H}_k^{(s)} - H_k^{(s)})^2}}{\sqrt{\sum_{k=1}^{N} (H_k^{(s)})^2}},
\end{equation}
where $\hat{H}^{(s)}_{k}$ denotes the DeepONet predicted value at spatial node $k$ for scenario $s$, ${H^{(s)}_{k}}$ denotes the corresponding ground-truth value simulated by the SWAN solver, and $N$ denotes the total number of spatial nodes in the computational domain. 

Moreover, the scenario-based mean absolute error (MAE) and scenario-based root mean square error (RMSE) are defined as: 
\begin{equation}
    MAE^{(s)} = \frac{1}{N} \sum_{k=1}^{N}\left| H_{k}^{(s)} - \hat{H}_{k}^{(s)} \right|,
\end{equation}
\begin{equation}
RMSE^{(s)} = \sqrt{\frac{1}{N} \sum_{k=1}^{N} \left( H_{k}^{(s)} - \hat{H}_{k}^{(s)} \right)^{2}}.
\end{equation}
Furthermore, we define the scenario-based maximum AE to capture the highest point-wise error for a given scenario $s$:
\begin{equation}
AE_{M}^{(s)} = \max_k \left|\hat{H}_k^{(s)} - H_k^{(s)}\right|.
\end{equation}
The subscript $M$ is adopted here to denote the maximum value within a specific scenario.

\subsection{Node-wise Spatial Metrics}
Although scenario-based metrics provide an overview of the model reliability, they do not resolve the spatial distribution of errors. To assess the surrogate's performance across the physical domain, the node-wise absolute error (AE) at node $k$ is defined as:
%
% Requires: \usepackage{amsmath}
\begin{equation}
AE_{k} = \left|\hat{H}_{k} - H_{k}\right|.
\end{equation}
%
%where $\hat{H}_{k}$ denotes the DeepONet predicted value at spatial node $k$ and ${H_{k}}$ denotes the corresponding ground-truth value simulated by the SWAN solver. 

In addition, the node-wise relative $L_2$ error at node $k$ for the $s$th scenario is computed as:
\begin{equation}
RLE_k^{(s)} = \frac{AE_k}{\| H^{(s)} \|_2} 
\end{equation}

 These node-wise error metrics reveal where the largest prediction discrepancies occur within the spatial domain, providing insight into whether the DeepONet struggles in particular regions (e.g., nearshore zones, boundaries, or around steep bathymetric gradients). \\

\section{Numerical Experiments and Results}\label{sec3}

\subsection{1-D Model Performance with 5\% Spatial Sampling Baseline}
\begin{figure}[h!]  
\centering
\includegraphics[width=0.8\textwidth,trim=0 0cm 0 0,clip]{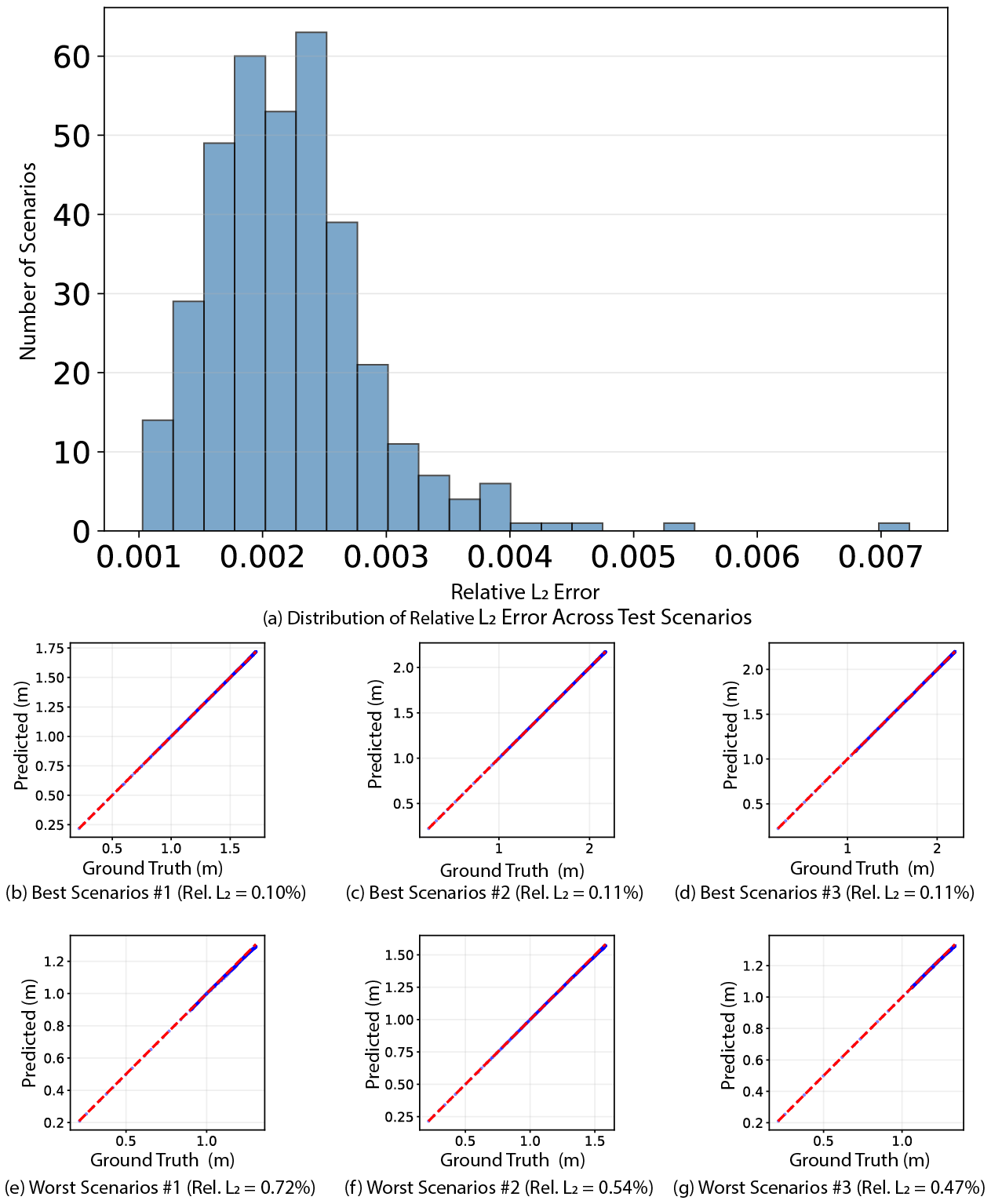}  
\caption{Analysis of Hsig prediction errors for the 1-D example. (a) Histogram of scenario-based relative $L_2$ errors across all test scenarios. (b-g) Parity plots comparing predicted and ground-truth values for the three best-performing scenarios (middle row) and the three worst-performing scenarios (bottom row), where the dashed diagonal lines indicate ideal agreement.}
\label{fig:1D_Hsig_Distribution} 
\end{figure}
\begin{figure}[h!]  
\centering
\includegraphics[width=0.8\textwidth,trim=0 0cm 0 0,clip]{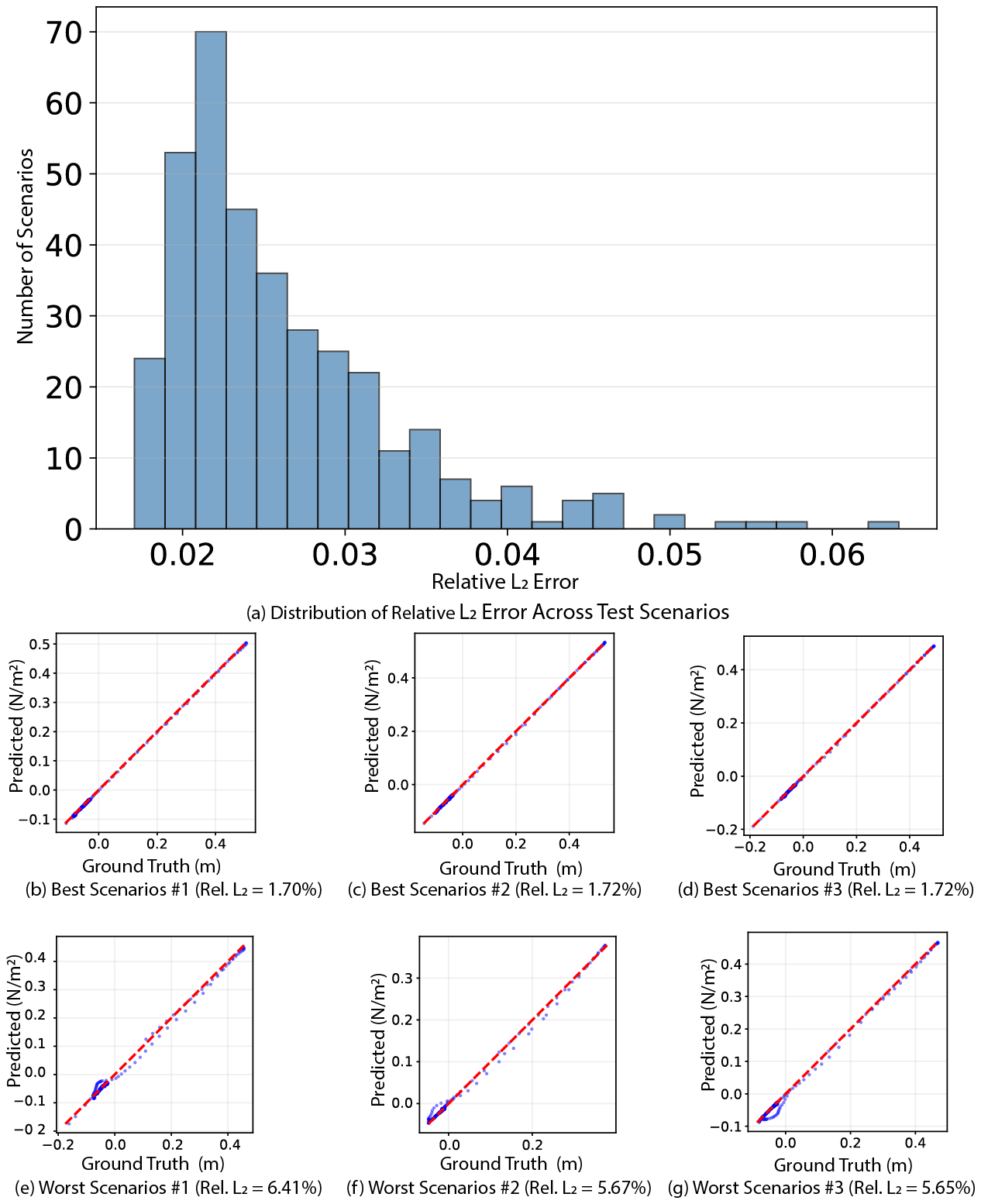}  
\caption{Analysis of radiation stress gradients prediction errors for the 1-D example. (a) Histogram of scenario-based relative $L_2$ errors across all test scenarios. (b-g) Parity plots comparing predicted and ground-truth values for the three best-performing scenarios (middle row) and the three worst-performing scenarios (bottom row), where the dashed diagonal lines indicate ideal agreement.}
\label{fig:1D_forces_Distribution} 
\end{figure}

Figure~\ref{fig:1D_Hsig_Distribution} and Figure~\ref{fig:1D_forces_Distribution} summarize the performance of the DeepONet models for Hsig and radiation stress gradients, respectively. The histograms illustrate the distribution of $RLE^{(s)}$ across the entire test set. For Hsig (Figure~\ref{fig:1D_Hsig_Distribution}(a)), the errors are tightly clustered between $0.1\%$ and $0.4\%$. This indicates that the surrogate performs consistently throughout the parameter space. Although the relative $L_2$ errors for the radiation stress gradients (Figure~\ref{fig:1D_forces_Distribution}(a)) are marginally higher in magnitude, the model achieves a $RLE^{(s)}$ below $5\%$ for the majority of scenarios. 

Based on the relative $L_2$ errors, three test scenarios with the lowest errors and three with the highest errors are chosen as the best and worst scenarios, respectively, for both Hsig and radiation stress gradients. As shown in the parity plots in Figures~\ref{fig:1D_Hsig_Distribution}(b-d) and Figures~\ref{fig:1D_forces_Distribution}(b-d), the best scenarios demonstrate near-perfect alignment between the true and predicted values for Hsig and radiation stress gradients, respectively. In the worst scenarios, Hsig predictions remain well-aligned (Figures~\ref{fig:1D_Hsig_Distribution}(e-g)); however, the radiation stress gradients exhibit a marginally higher degree of deviation (Figures~\ref{fig:1D_forces_Distribution}(e-g)). 
% This discrepancy is attributed to the inherent sensitivity of spatial derivatives of forces, especially in regions where the force magnitude transitions.

The three worst scenarios of each variable are selected to compute their Max AE, MAE, and RMSE. The corresponding performance metrics and scenario details are presented in Table~\ref{tab:1d_worst_cases_magnitudes} and Table~\ref{tab:1d_worst_cases_scenarios}, respectively. 

\begin{figure}[h!]
\centering
\includegraphics[width=0.9\textwidth]{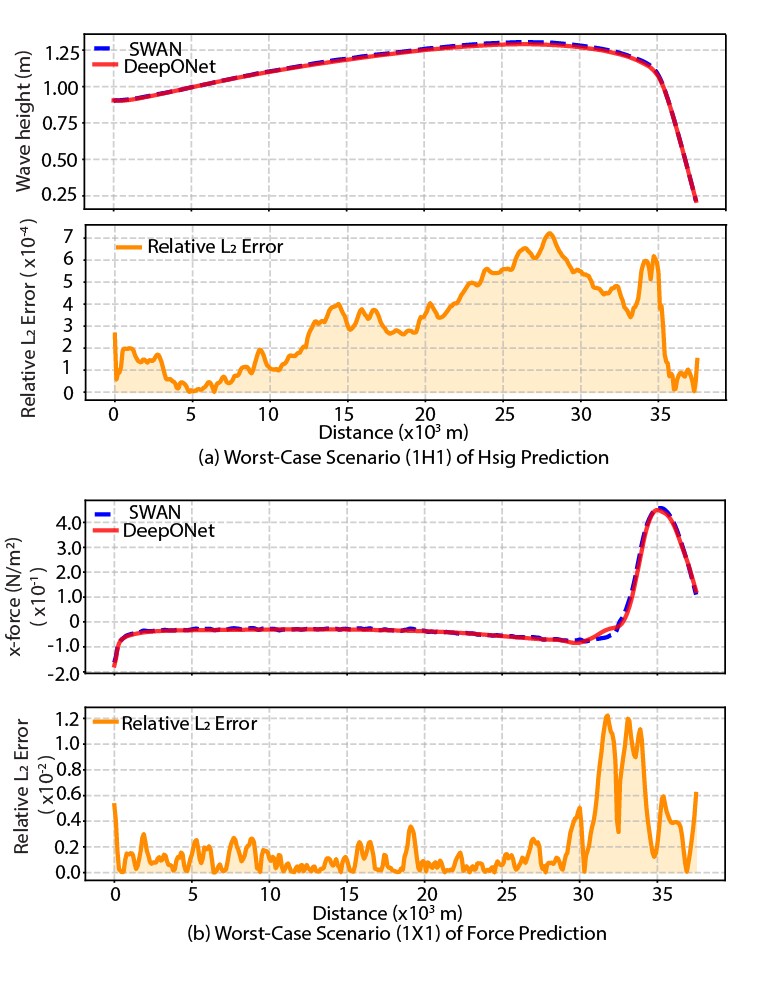}
\caption{Comparison between SWAN and DeepONet for the worst-case scenarios in the 1-D example. The top panel shows the Hsig prediction and the bottom panel shows the radiation stress gradient predictions.}
\label{fig:1d_oneScenario}
\end{figure}

\begin{table*}[htb]
\centering
\begin{tabular}{cccccc}
\textbf{Variable} & \textbf{Scenario\footnotemark} & $\mathbf{RLE^{(s)} (\%)}$ & 
$\mathbf{AE_M^{(s)}}$ & $\mathbf{MAE^{(s)}}$ & $\mathbf{RMSE^{(s)}}$ \\ \midrule
\textbf{Hsig}
& 1H1
& 0.72
& $1.61\times10^{-2}$ 
& $7.00\times10^{-3}$ 
& $8.31\times10^{-3}$ \\
& 1H2
& 0.54 
& $1.54\times10^{-2}$ 
& $5.50\times10^{-3}$ 
& $7.13\times10^{-3}$ \\
& 1H3
& 0.47  
& $1.11\times10^{-2}$ 
& $4.40\times10^{-3}$ 
& $5.71\times10^{-3}$ \\ \midrule
% ---------------- Forces -----------------
\textbf{Forces}
& 1X1
& 6.41
& $2.98\times10^{-2}$ 
& $4.85\times10^{-3}$ 
& $8.03\times10^{-3}$ \\
& 1X2
& 5.67
& $2.47\times10^{-2}$ 
& $2.58\times10^{-3}$ 
& $5.07\times10^{-3}$ \\
& 1X3
& 5.65 
& $3.66\times10^{-2}$ 
& $3.47\times10^{-3}$ 
& $7.13\times10^{-3}$ \\ 
\end{tabular}
\caption{Error metrics of the three worst-performing scenarios for each variable in the 1-D example, identified by the three highest $RLE^{(s)}$ magnitudes across all test cases.}
\label{tab:1d_worst_cases_magnitudes}
\end{table*}
\footnotetext{
Scenarios are labeled using the format $N_1 V N_2$, where:
$N_1$ represents the example (1: 1-D, 2: DUCK, C: Cross-Grid),
$V$ represents the variable (H: Hsig, X: X-forces, Y: Y-forces), and
$N_2$ represents the severity rank (1: worst, 2: second worst, 3: third worst).
For example, ``1H3'' represents the scenario with the third highest relative $L_2$ error in predicting Hsig for the 1-D example.
}
\begin{table}[htb]
\centering
\begin{tabular}{cccccc}
\textbf{Variable} & \textbf{Scenario} & 
\textbf{Wind} & \textbf{Wind Dir} &
\textbf{Wave H.} & \textbf{Wave Dir}  \\ \midrule
\textbf{Hsig}
& 1H1
& 11 & -7.0 & 0.9 & 6.0  \\
& 1H2   
& 13 & -4.0 & 0.6 & 0.0   \\
& 1H3 
& 11 & -7.0 & 1.2 & 6.0  \\ \midrule
% ---------------- force -----------------
\textbf{Forces}
& 1X1
& 15 & -7.0 & 0.6 & 7.0 \\
& 1X2     
& 12 & 0.0  & 1.2   & 0.0  \\
& 1X3   
& 15 & -6.0  & 1.0   & 7.0   \\

\end{tabular}
\caption{Configurations of the three worst-performing scenarios reported in Table~\ref{tab:1d_worst_cases_magnitudes}. Wind speed is given in m/s, wave height in m, and directional quantities in degrees.}
\label{tab:1d_worst_cases_scenarios}
\end{table}
To further assess model robustness, we next examine the worst-performing scenarios for each variable, i.e. 1H1 and 1X1 (see Table~\ref{tab:1d_worst_cases_magnitudes}). Figure \ref{fig:1d_oneScenario} compares the model predictions for the Hsig and the radiation stress gradients with the SWAN simulation results. 
For Hsig, the worst scenario occurs for a boundary wave height of 0.9 m with a direction of $6^\circ$, and a wind speed of 11 m/s with a direction of $-7^\circ$ (see Table~\ref{tab:1d_worst_cases_scenarios}). The DeepONet predictions are closely aligned with the SWAN results (see Figure \ref{fig:1d_oneScenario}(a)), and the highest magnitude of the node-wise relative $L_2$ error ($RLE_k$) is $0.07\%$, which occurs at $x=2.8 \times 10^4$~m with a Hsig of $1.30$~m, as the wave begins to decay. 

For radiation stress gradients, the worst scenario involves an boundary wave height of $0.6$ m with a direction of $7^\circ$, and a wind speed of $15$ m/s with a direction of $-7^\circ$ (see Table~\ref{tab:1d_worst_cases_scenarios}). The highest $RLE_k$ of $1.22\%$ occurs at $x= 3.18 \times 10^4 m$, which corresponds to a point of sharp increase in the  radiation stress gradients as the wave approaches the nearshore environment, and where the true value simulated by SWAN is $-5.98\times10^{-2}~N/m^2$. As illustrated in Figure \ref{fig:1d_oneScenario}(b), the SWAN output exhibits a localized, sharp gradient, but the DeepONet model produces a smoother transition. Despite this tendency to regularize sharp gradients in the predicted field, Figures~\ref{fig:1D_forces_Distribution} and Figures~\ref{fig:1d_oneScenario} suggest that the DeepONet surrogate for the  radiation stress gradients captures the general trend of the solution and achieves a high degree of accuracy in the majority of the spatial domain, even for the worst test scenarios.

\subsection{Spatial Sampling Size Sensitivity and Trade-off Analysis}
To evaluate the impact of spatial sampling size on predictive accuracy and computational training overhead, a sensitivity analysis is conducted on the 1-D example. All models across the sampling sizes are trained using the same set of optimal hyperparameters determined via the Optuna optimization framework (Table~\ref{tab:best_hyperparams}).
\begin{figure}[h!]  
\centering
\includegraphics[width=1.0\textwidth]{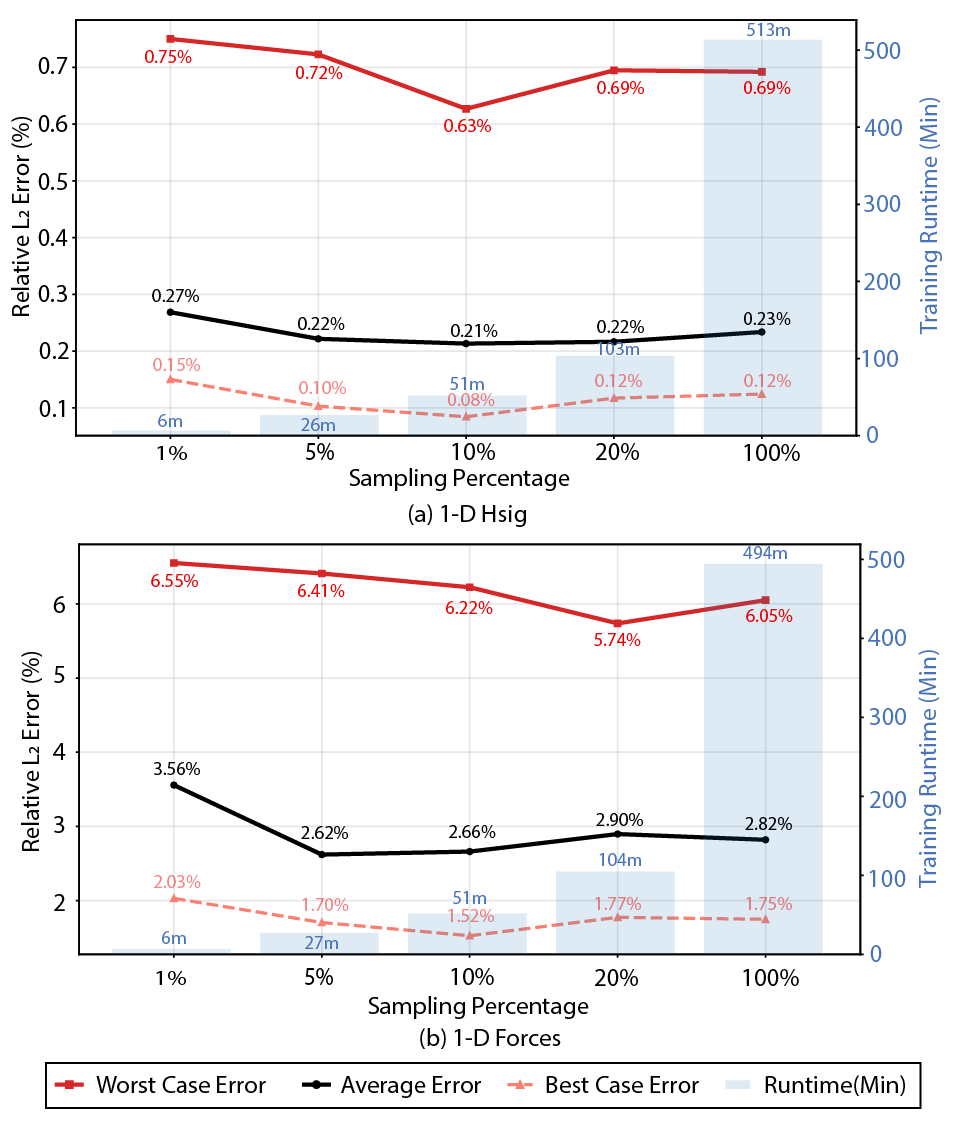}  
\caption{Sensitivity of model prediction (best-case, mean, and worst-case $RLE^{(s)}$) and computational training overhead to varying spatial coordinate sampling percentages ($1\%$, $5\%$, $10\%$, $20\%$, and $100\%$) over the idealized 1-D uniform slope domain: (a) evaluation for the $H_{sig}$, and (b) evaluation for the wave-induced radiation-stress force.}
\label{fig:SampleSizeSensitivity} 
\end{figure}
Figure~\ref{fig:SampleSizeSensitivity} illustrates the best-case, mean, and worst-case $RLE^{(s)}$ across all test scenarios, mapped against the corresponding training execution runtimes. For both Hsig and radiation stress gradients models, a huge accuracy improvement is observed when increasing the sampling ratio from 1\% to 5\%. For the mean errors, the 5\% sampling size is the optimal configuration for forces. When expanding the sampling from 5\% to 10\% for the Hsig model, the mean error improves by only 0.01\%, while training time doubles. However, worst-case evaluations indicate that it requires a 10\% sampling allocation for the Hsig and a 20\% allocation for the radiation stress gradients model. The best-case reaches its minimum at a 10\% sampling size.

A critical nuance of this sensitivity behavior relates to the computational constraints of the hyperparameter optimization process. Executing an automated Optuna is heavily resource-intensive due to the exploration of the multi-dimensional search space. Because the optimal hyperparameter combination used in this analysis (Table~\ref{tab:best_hyperparams}) was obtained using a fixed 5\% sampling size, a structural optimization bias is introduced. The resulting branch and trunk network layouts are optimized precisely for that specific data size. Consequently, when the spatial data is increased to 20\% without a corresponding expansion in network size, the accuracy degrades. This indicates that the fixed DeepONet model capacity becomes too constrained to learn and generalize across the larger set of spatial coordinate dimensions during training. Therefore, the observed performance at 5\% sampling is specific to this fixed hyperparameter configuration. Re-optimizing the network hyperparameters for higher sampling ratios might likely yield further performance improvements. 

In summary, due to the marginal improvement in predictive skill and the significant increase in computational cost observed by increasing the spatial sampling ratio above 5\%, expensive sets of comprehensive Optuna hyperparameter optimization studies are not conducted for higher sampling ratios, and the 5\% spatial sampling ratio is selected for the rest of this work. 

\subsection{DUCK Example}
\begin{figure}[h!]  
\centering
\includegraphics[width=0.9\textwidth]{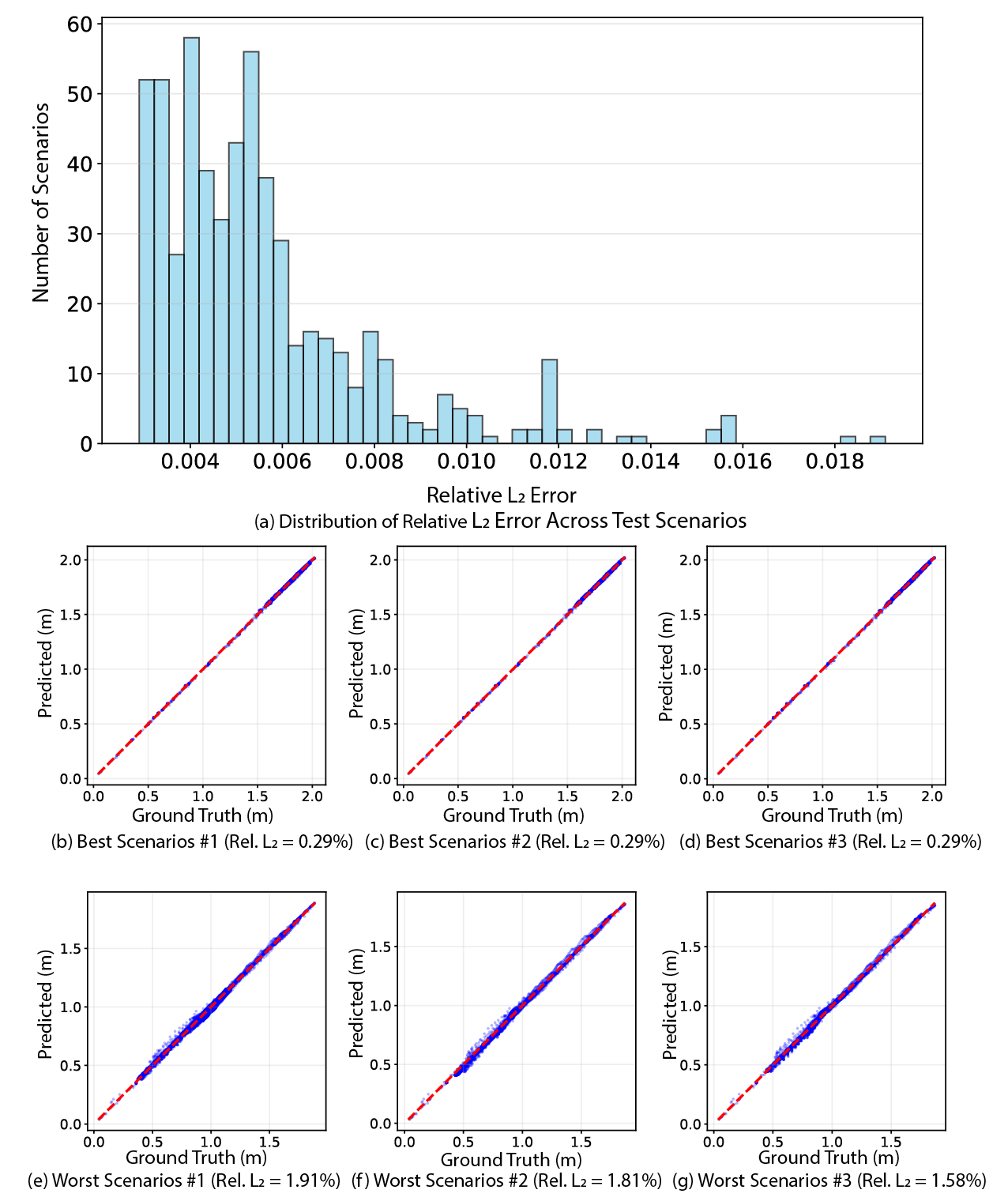}  
\caption{Analysis of Hsig prediction errors for the DUCK test example. (a) Histogram of scenario-based relative $L_2$ errors across all test scenarios. (b-g) Parity plots comparing predicted and ground-truth values for the three best-performing scenarios (middle row) and the three worst-performing scenarios (bottom row), where the dashed diagonal lines indicate ideal agreement.}
\label{fig:duck_Hsig_Distribution} 
\end{figure}

\begin{figure}[h!]  
\centering
\includegraphics[width=0.9\textwidth]{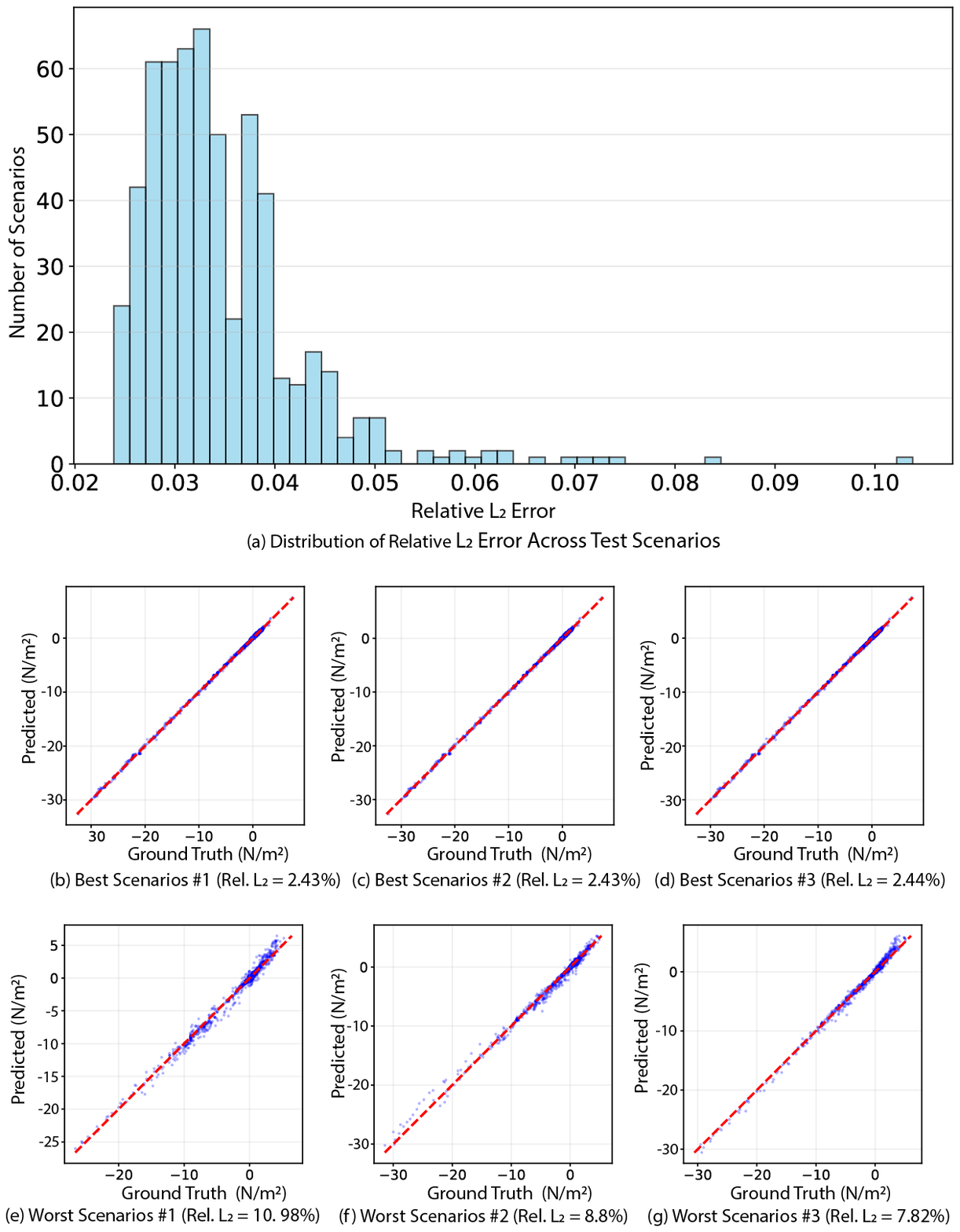}  
\caption{Analysis of x-forces prediction errors for the DUCK test example. (a) Histogram of scenario-based relative $L_2$ errors across all test scenarios. (b-g) Parity plots comparing predicted and ground-truth values for the three best-performing scenarios (middle row) and the three worst-performing scenarios (bottom row), where the dashed diagonal lines indicate ideal agreement.}
\label{fig:duck_xforces_Distribution} 
\end{figure}

\begin{figure}[h!]  
\centering
\includegraphics[width=0.9\textwidth]{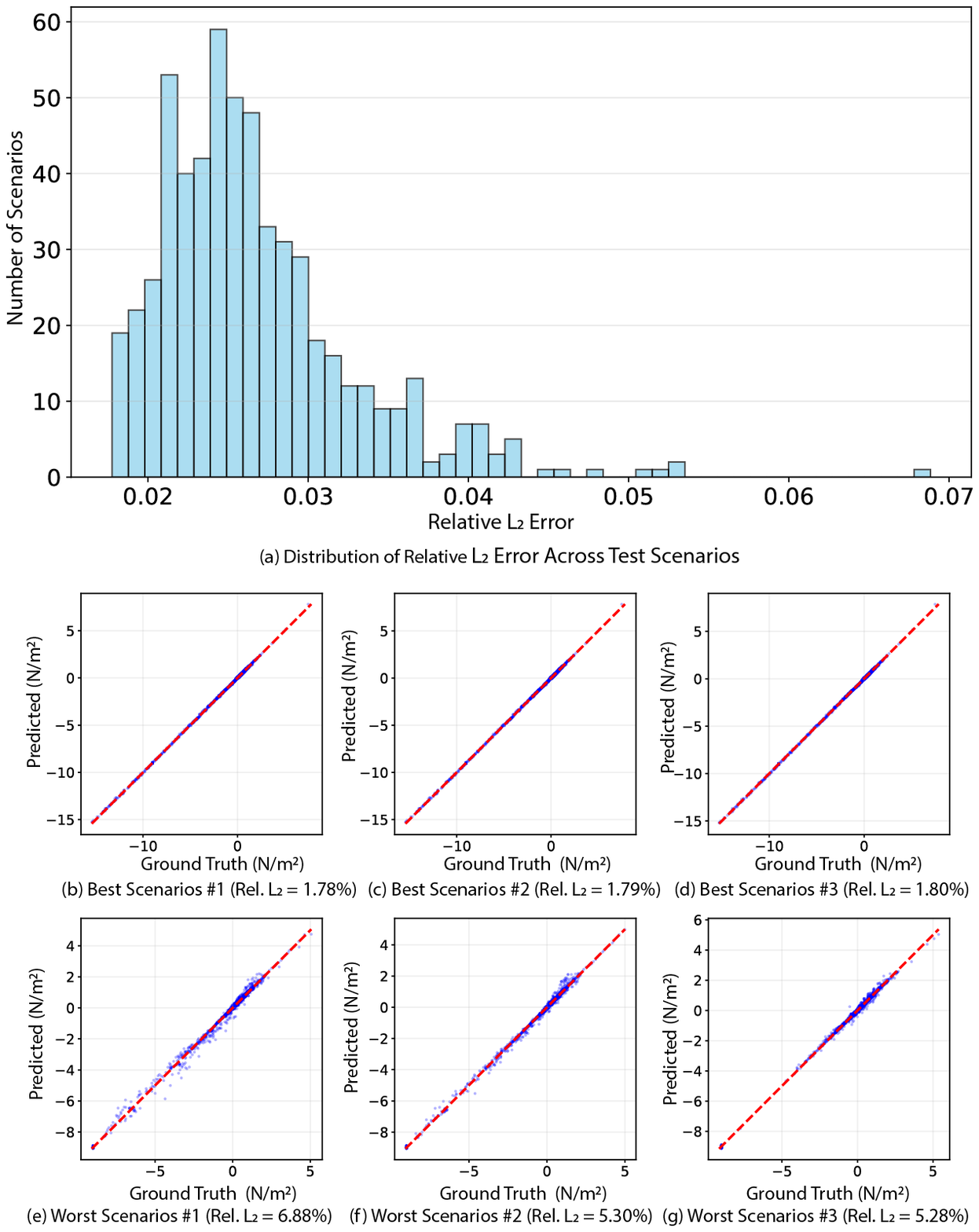}  
\caption{Analysis of y-forces prediction errors for the DUCK test example. (a) Histogram of scenario-based relative $L_2$ errors across all test scenarios. (b-g) Parity plots comparing predicted and ground-truth values for the three best-performing scenarios (middle row) and the three worst-performing scenarios (bottom row), where the dashed diagonal lines indicate ideal agreement. }
\label{fig:duck_yforces_Distribution} 
\end{figure}

The performance analyses of the DeepONet surrogate models for Hsig and the x- and y-forces are presented in Figures~\ref{fig:duck_Hsig_Distribution}, \ref{fig:duck_xforces_Distribution} and  \ref{fig:duck_yforces_Distribution}, respectively. Despite the highly irregular seabed features, the $RLE^{(s)}$ distributions confirm that the models maintain strong global consistency. For Hsig predictions, the $RLE^{(s)}$ values (see Figure~\ref{fig:duck_Hsig_Distribution}(a)) are below $1.0\%$ for the majority of test scenarios. Although this is a minor increase compared to the idealized 1-D example, it remains remarkably low for a real-world field application. The $RLE^{(s)}$ for the $x$-direction forces (see Figure~\ref{fig:duck_xforces_Distribution}(a)) typically range between $2\%$ and $5\%$, with a maximum value of $10.98\%$. The $y$-direction forces exhibit a similar distribution (see Figure~\ref{fig:duck_yforces_Distribution}(a)), with the majority of errors less than $3.5\%$ and a maximum of $6.88\%$.

Analysis of the parity plots reveals that the best-performing scenarios for all variables (Figures~\ref{fig:duck_Hsig_Distribution}(b-d), \ref{fig:duck_xforces_Distribution}(b-d) and  \ref{fig:duck_yforces_Distribution}(b-d)) show that the predicted solutions are nearly indistinguishable from the true solutions. However, the relatively higher levels of mismatch between the true and predicted solutions, seen in the parity plots of the worst-performing scenarios (Figures~\ref{fig:duck_Hsig_Distribution}(e-g), \ref{fig:duck_xforces_Distribution}(e-g) and  \ref{fig:duck_yforces_Distribution}(e-g)) highlight the challenges posed by real-world bathymetry, as well as numerical artifacts introduced near the boundary. For example, the minimum value of the x-force solution for scenario 3X1 is approximately $-67~N/m^2$, which occurs at a single isolated grid point on the southern boundary, whereas the next highest solution value at a grid point is $-33~N/m^2$. This isolated point, which is an outlier and potentially an effect of convergence issues of the numerical solution near the boundary, is excluded from the parity plot in Figure~\ref{fig:duck_xforces_Distribution}(e). Similarly to the previous two examples, the Hsig model demonstrates impressive performance even in the worst scenarios. The model successfully captures the overall energy dissipation across the complex physical system of the DUCK example. The force predictions in the worst scenarios exhibit much more scattering than in the 1-D cases. This increased misalignment is attributed to the much wider range of unseen flow conditions and the presence of irregular, realistic bathymetry features, which create not only a higher contrast in the magnitude of the forces, but also localized, high-frequency spatial gradients. These effects are discussed in the following analysis.

To further assess the robustness of the models, the three worst-performing test scenarios for each variable are selected to compute their $AE_M^M$, $MAE^M$, and $RMSE^M$. The corresponding error metrics and the scenario configurations are presented in 
Tables~\ref{tab:duck_worst_cases_magnitude} and~\ref{tab:duck_worst_cases}, respectively. 
Given the increased physical and geometric complexity of the DUCK example, the scenarios associated with the largest $RLE^{(s)}$ for each variable, i.e., 2H1, 2X1, and 2Y1, are selected for detailed spatial comparison between SWAN and DeepONet. 
\begin{table*}[h]
\centering
\begin{tabular}{cccccc}
\textbf{Variable} & \textbf{Scenario} & $\mathbf{RLE^{(s)} (\%)}$ & 
$\mathbf{AE_M^{(s)}}$ & $\mathbf{MAE^{(s)}}$ & $\mathbf{RMSE^{(s)}}$ \\\midrule
\textbf{Hsig}
& 2H1
& 1.91
& $9.87\times10^{-2}$ 
& $1.44\times10^{-2}$ 
& $1.86\times10^{-2}$ \\
& 2H2
& 1.81   
& $9.54\times10^{-2}$ 
& $1.29\times10^{-2}$ 
& $1.75\times10^{-2}$ \\
& 2H3
& 1.58
& $1.00\times10^{-1}$ 
& $1.15\times10^{-2}$ 
& $1.53\times10^{-2}$ \\\midrule
\textbf{x-forces}
& 2X1
& 10.98
& $2.55$ 
& $1.30\times10^{-1}$ 
& $2.54\times10^{-1}$ \\
& 2X2
& 8.8   
& $2.75$ 
& $9.81\times10^{-2}$ 
& $2.12\times10^{-1}$ \\
& 2X3
& 7.8  
& $2.51$ 
& $8.86\times10^{-2}$ 
& $1.85\times10^{-1}$ \\\midrule
\textbf{y-forces}
& 2Y1
& 6.88
& $1.51$ 
& $3.52\times10^{-2}$ 
& $8.69\times10^{-2}$ \\
& 2Y2
& 5.30   
& $9.13\times10^{-1}$ 
& $3.15\times10^{-2}$ 
& $6.68\times10^{-2}$ \\
& 2Y3
& 5.28  
& $7.68\times10^{-1}$ 
& $2.87\times10^{-2}$ 
& $5.44\times10^{-2}$ \\

\end{tabular}
\caption{Error metrics of the three worst-performing scenarios for each variable in the Duck example, identified by the three highest $RLE^{(s)}$ magnitudes across all test cases.}
\label{tab:duck_worst_cases_magnitude}
\end{table*}
\begin{table*}[h]
\centering
\begin{tabular}{cccccc}
\textbf{Variable} & \textbf{Scenario} & 
\textbf{Wind} & \textbf{Wind Dir} &
\textbf{Wave H.} & \textbf{Wave Dir}  \\\midrule
\textbf{Hsig}
& 2H1
& 7 & 315 & 1.6 & -20  \\
& 2H2
& 7 & 90 & 1.6 & -20  \\
& 2H3  
& 6 & 135 & 1.6 & -20 \\\midrule
\textbf{x-forces}
& 2X1
& 8 & 0 & 1.6 & 160 \\
& 2X2 
& 7 & 90 & 1.6 & -20 \\
& 2X3  
& 6 & 135 & 1.6 & -20  \\\midrule
\textbf{y-forces}
& 2Y1
& 7 & 90 & 1.6 & -20  \\
& 2Y2   
& 6 & 135 & 1.6 & -20  \\
& 2Y3  
& 8 & 225 & 1.6 & 160  \\
\end{tabular}
\caption{Configurations of the three worst-performing scenarios for each variable in DUCK example, as reported in Table~\ref{tab:duck_worst_cases_magnitude}. Wind speed is given in m/s, wave height in m, and directional quantities in degrees. }

\label{tab:duck_worst_cases}
\end{table*}
\clearpage
\begin{figure}[h!]  
\centering
\includegraphics[width=0.9\textwidth]{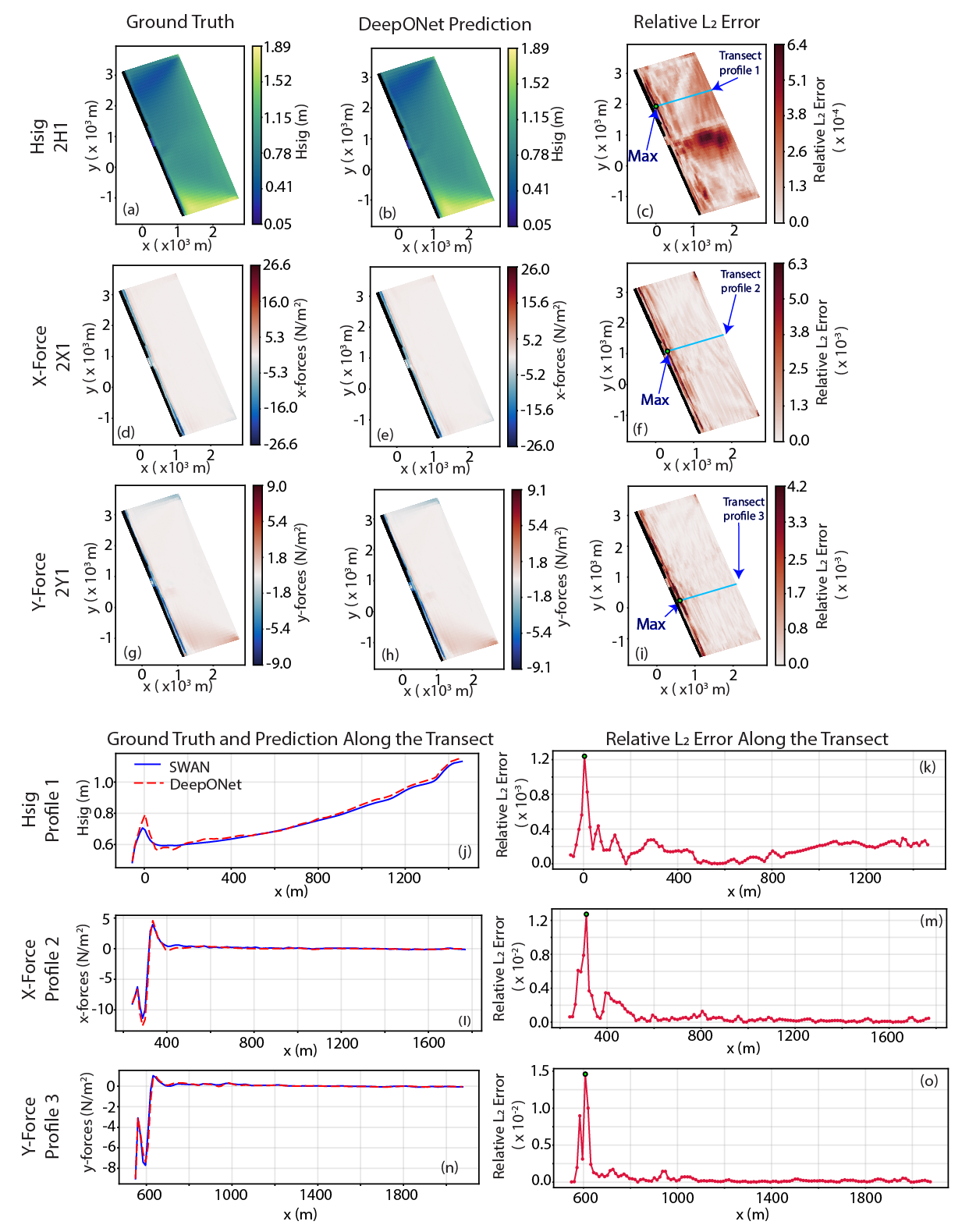}  
\caption{Comparison of the worst-case scenarios for each variable between SWAN and DeepONet in the DUCK example. Wind and wave settings are specified in Table~\ref{tab:duck_worst_cases}. Panels (a–c) show the ground truth, DeepONet prediction, and relative $L_2$ error for the worst Hsig scenario. Panels (d–f) and Panels (g–i) present the ground truth, DeepONet prediction, and relative $L_2$ error for the worst x-force and y-force predictions, respectively. For enhanced spatial visualization, the color bar in (c), (f), and (i) is clipped at the 99th percentile to reduce the influence of extreme values and better highlight the overall spatial error patterns. Finally, panels (j-k), (l-m) and (n-o) show the cross-shore profiles and RLE of Hsig, x-forces, and y-forces, respectively, along the transects shown in (c), (f) and (i), respectively. }
\label{fig:duck_scenario} 
\end{figure}
\clearpage
\begin{figure}[h!]  
\centering
\includegraphics[width=0.9\textwidth]{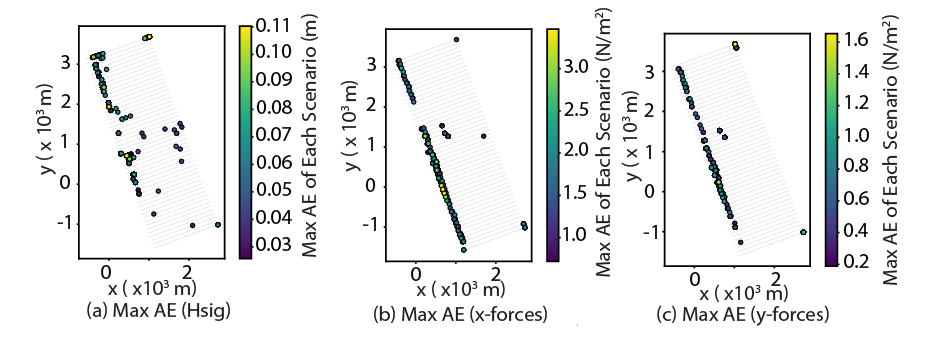}  
\caption{Location of the maximum absolute error for every test scenario of the DUCK example.}
\label{fig:duck_mae} 
\end{figure}
As shown in Figure \ref{fig:duck_scenario}, the maximum $RLE_k$ for the Hsig in scenario 2H1 (first row) is 0.064\%, occurring at a true wave height of $0.693$ m. For the x-direction force comparison in scenario 2X1 (second row), the maximum $RLE_k$ is $0.62\%$, which occurs at a force of $-3.25$ $N/m^2$. As for the y-direction force prediction in scenario 2Y1 (third row), the maximum $RLE_k$ is $0.41\%$, which occurs at a force of $-4.35$ $N/m^2$. Across all three predicted variables, the maximum $RLE_k$ tends to occur either in the nearshore zone, where wave transformation processes such as shoaling and breaking induce strong spatial gradients, or along the offshore boundary. This is further illustrated in Figure~\ref{fig:duck_mae}, which shows the maximum absolute error for each scenario in the DUCK test set. The results indicate that the largest discrepancies are predominantly concentrated in the nearshore region, where the complex bathymetry and wave-breaking physics introduce significant non-linearities.

The results for the Duck example demonstrate that the DeepONet surrogate is capable of learning the underlying wave physics of a complex, real-world coastal environment. Although irregular bathymetry increases the difficulty of predicting derivative-based quantities, such as forces, the global relative $L_2$ errors remain well within acceptable thresholds for practical coastal engineering and coupled hydrodynamic modeling.

\subsection{Comparative Analysis against Baseline Architectures}

To validate the structural advantages of the proposed DeepONet, it is compared to a standard Multi-Layer Perceptron (MLP) baseline architecture using the DUCK example. The MLP baseline uses the Branch Network input as its input feature and directly predicts the bulk wave parameter fields on the entire SWAN computational mesh. Thus, The MLP output is a $6703$-dimensional vector corresponding to the discrete ground-truth labels on the SWAN numerical mesh.

Two distinct variations of the MLP baseline are evaluated. The first configuration, denoted MLP-2L, utilizes a compact two-layer hidden architecture with widths of 4 and 8 neurons, yielding a total trainable parameter budget comparable to that of the DeepONet models. The second baseline, denoted MLP-3L, is a deeper, scaled-up model with three hidden layers containing 32, 64, and 32 neurons. Table~\ref{tab:trainableParams} shows the total number of trainable parameters across all tested architectures. 
\begin{table*}[h]
\centering
\begin{tabular}{cccc}
\textbf{Variable} & DeepONet & MLP-2L & MLP-3L\\
\midrule
\textbf{Hsig} & 40,729 & 60,397 & 225,583 \\
\midrule
\textbf{x-forces} & 94,621 & 60,397 & 225,583  \\
\midrule
\textbf{y-forces} & 94,621 & 60,397  & 225,583  \\
\end{tabular}
\caption{Total trainable parameters of the DeepONet and MLP benchmark models for each variable in the DUCK example.}
\label{tab:trainableParams}
\end{table*}

\begin{figure}[h!]  
\centering
\includegraphics[width=0.9\textwidth]{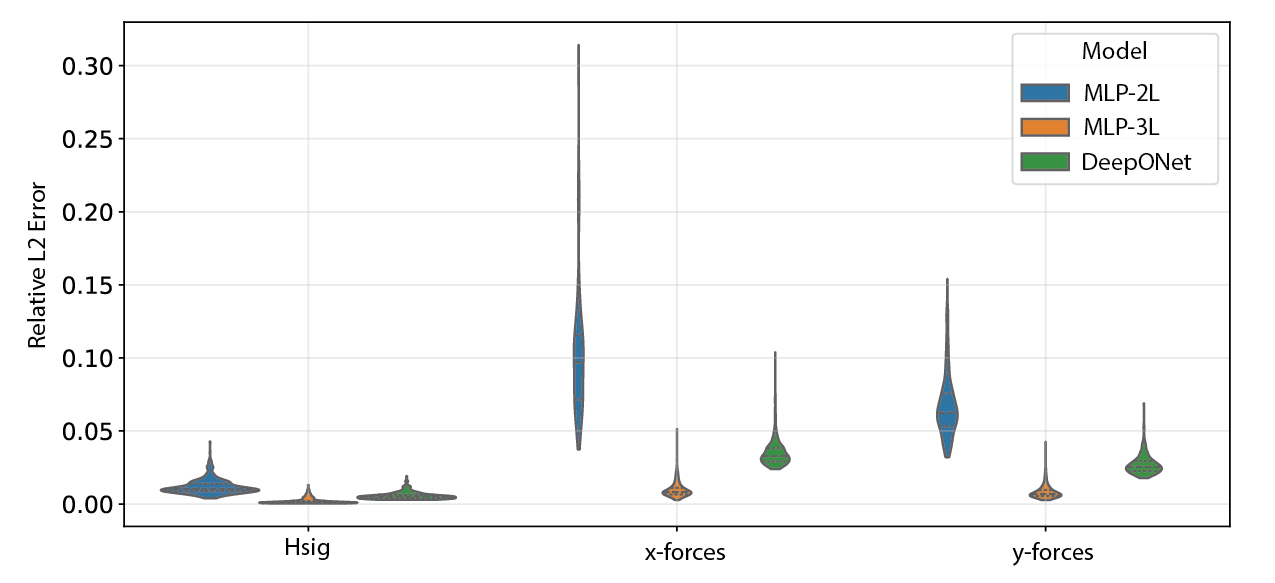}  
\caption{Distribution of test scenario $RLE^{s}$ values of the DeepONet and MLP models for each variable}
\label{fig:mlp_3models} 
\end{figure}

Table~\ref{tab:mlp_ModelPerformance} and Figure~\ref{fig:mlp_3models} present the quantitative performance comparisons across all 576 DUCK testing scenarios. Table~\ref{tab:mlp_ModelPerformance} presents the mean, median and worst $RLE^{s}$ values for each variable across the test scenarios. Figure~\ref{fig:mlp_3models} shows the distribution of test scenario $RLE^{s}$ values of the DeepONet and the two MLP models for all three modeled variables. The results reveal that, among models with similar parameter capacity, the proposed DeepONet outperforms the MLP-2L baseline across all physical variables. On the other hand, the MLP-3L model achieves highly competitive local interpolation metrics on the training grid. However, its output dimensions are strictly tied to the initial spatial discretizations; it cannot handle variable-mesh configurations. 

It is important to note that the comparison favors MLP architectures in terms of data availability. Both MLP models are trained using the complete spatial field on the original grid, thereby  learning from the solution labels at all available coordinate locations. In contrast, the proposed DeepONet is trained using only 5\% randomly sampled spatial coordinates. Despite being trained on substantially fewer spatial observations, DeepONet consistently achieves superior predictive performance while retaining the ability to generalize across arbitrary discretizations. 
Therefore, this comparison study also indirectly illustrates the discretization-invariance property of the DeepONet architecture; it learns the underlying continuous mathematical operator by processing continuous coordinates independently through its trunk network.

\begin{table*}[h]
\centering
\begin{tabular}{ccccc}
\textbf{Variable} & \textbf{Metrics} & 
\textbf{DeepONet} & \textbf{MLP-2L} &
\textbf{MLP-3L}  \\\midrule
\textbf{Hsig}
& Mean $RLE^{(s)}$
& 0.56\% & 1.17\% & 0.20\%  \\
& Median $RLE^{(s)}$  
& 0.51\% & 1.03\% & 0.10\%   \\
& Worst $RLE^{(s)}$  
& 1.91\% & 4.27\% & 1.31\% \\\midrule
\textbf{x-forces}
& Mean $RLE^{(s)}$
& 3.46\% & 10.18\% & 0.98\% \\
& Median $RLE^{(s)}$
& 3.27\% & 9.71\% & 0.82\%  \\
& Worst $RLE^{(s)}$
& 10.38\% & 31.42\% & 5.13\%  \\\midrule
\textbf{y-forces}
& Mean $RLE^{(s)}$
& 2.67\% & 6.62\% & 0.85\%   \\
& Median $RLE^{(s)}$
& 2.56\% & 6.24\% & 0.69\%   \\
& Worst $RLE^{(s)}$
& 6.89\% & 15.4\% & 4.26\%   \\
\end{tabular}
\caption{Error metrics for the DeepONet and MLP models across all test scenarios of the DUCK example}
\label{tab:mlp_ModelPerformance}
\end{table*}

\subsection{Discretization Invariance and Cross-Grid Verification}
% A significant advantage of DeepONet over traditional deep neural networks (DNNs) is its property of discretization invariance. While the input and output structures of traditional DNN architectures are fixed by the discrete representations of the input features and target functions seen during training, DeepONet treats the spatial coordinates of the output function as continuous-domain inputs dynamically queried through its trunk network, which can vary during training and inference. As noted previously, the DeepONet models in this work are trained on 5\% of all the available spatial data points, obtained by random sampling for every training scenario. Despite being trained on sparsely subsampled data, the ability to accurately infer the solution fields on the entire SWAN computational mesh is a direct consequence of the discretization invariance property of the architecture. 

It has been demonstrated that for both the 1D and 2D numerical examples, DeepONet surrogate models trained on $5\%$ randomly sampled spatial points can predict a spatially continuous solution field on the full grid, and can outperform MLP-based deep neural networks of similar capacity even when they are trained on the full grid solution. These results illustrate the mesh-independent prediction capabilities and specifically the cross-grid generalizability of the DeepONet models. To further evaluate the discretization invariance of the trained surrogate on completely unseen coordinates, a cross-grid verification study is presented using the DUCK example. 
% It should be emphasized that the primary objective of this cross-grid verification is to demonstrate DeepONet's discretization-invariant generalization. 

As a purely data-driven operator network, DeepONet learns the underlying physics embedded within the training dataset, and thus inherits the computational characteristics and limitations of the target numerical solver, including potential grid sensitivities of derived quantities of interest and under-resolved physical processes on coarse meshes. Thus, this evaluation serves primarily as a proof of concept, demonstrating that DeepONet can evaluate modeled variables at unobserved spatial coordinates without retraining. For practical engineering applications and validation against real-world measurements, the numerical fidelity and convergence of the training dataset must be thoroughly established first. 

\begin{figure}[h!]  
\centering
\includegraphics[width=0.9\textwidth]{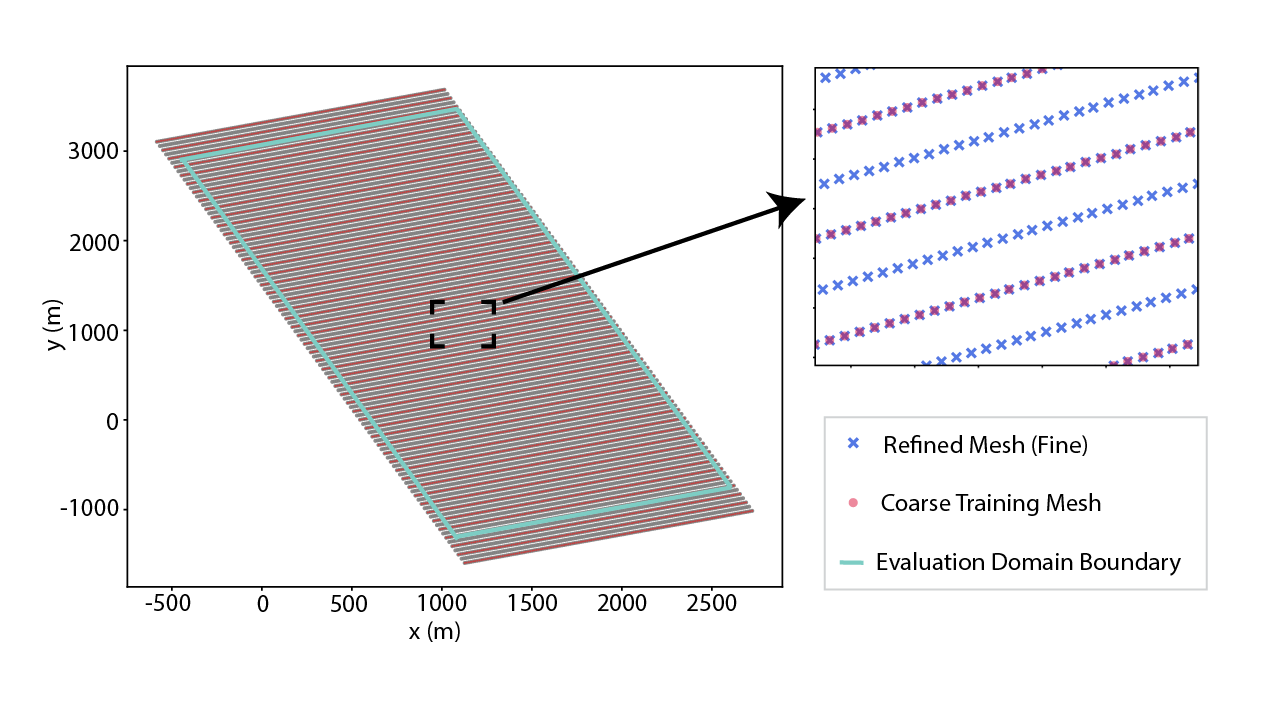}  
\caption{Coarse and fine meshes used in the discretization invariance study. The region used in the evaluation of the error metrics, after applying boundary truncation, is shaded in light blue color.}
\label{fig:RefindementGrid} 
\end{figure}

In this study, the DeepONet model, trained using 5\% random sampling of the bulk parameter values on the original coarse SWAN mesh, is used to evaluate the modeled variables on a refined mesh, and the model predictions are compared with the SWAN solutions simulated using the refined mesh. Figure~\ref{fig:RefindementGrid} shows the refined mesh layout overlaid on the original coarse mesh. This refined mesh is constructed by uniformly splitting every along-shore grid cell into two, i.e., the original coarse mesh has $137 \times 51 = 6987$ nodes (cross shore $\times$ along shore) whereas the refined mesh has $137 \times (2 \cdot 51 - 1) = 13837$ nodes. 

First, the mesh-size sensitivities of the computed bulk parameters of interest, namely, Hsig and the directional radiation stress gradients, are evaluated by comparing their values obtained from the high-resolution SWAN solution to those obtained from the baseline coarse-grid SWAN solution. Variable values are compared at grid points that are common between the coarse and fine mesh. Hsig remains highly robust, with a mean $RLE^{(s)}$ of only 0.93\% across all test scenarios. This is a physically consistent behavior, since Hsig depends on integrated source and sink terms, such as wind input and bottom friction, which smooth out rough spatial irregularities, thus reducing mesh sensitivity. However, the radiation stress gradient is found to be more sensitive to the mesh resolution, yielding mean $RLE^{(s)}$ of 6.5\% for x-direction and 18.11\% for y-direction stress gradients, as shown in Table~\ref{tab:ModelPerformance} (3rd column). This increased sensitivity can be primarily attributed to the fact that radiation stress gradients are computed by taking spatial derivatives of the radiation stress tensor (see eq.~\ref{eq:forces_definition}), which leads to amplification of any moderate to large rates of change of wave height (such as those caused by sandbars) or bathymetry (such as fine-scale irregularities in the seabed). These can often induce rapid fluctuations in the radiation stress gradients, which are typically under-resolved by first-order backward space or upwind schemes, adopted by numerical solvers such as SWAN
. Consequently, numerical diffusion or directional bias of upwinding may introduce non-physical discrepancies between the coarse and fine mesh simulations. Moreover, to ensure that the fine-grid results represent a true physical refinement rather than introducing spurious numerical artifacts, interpolation of the bathymetry field to the finer grid needs to be done precisely. On the other hand, the specification of the boundary JONSWAP spectrum needs to be verified to ensure that the total energy flux entering the domain is consistent across both spatial resolutions. Additionally, the resolution of the wave spectrum needs to be refined proportionately to mitigate potential inaccuracies due to the Garden Sprinkler Effect (GSE), a numerical artifact in which discrete spectral propagation causes a continuous wave field to unrealistically fragment into isolated energy patches, especially if SWAN's default GSE mitigation diffusion is insufficient to minimize GSE artifacts. 

As the DUCK example inherits the computational grid and other details from the F71 field experiment case study of the ONR Test Bed, a comprehensive grid convergence study to completely address these potential sources of discrepancies is well beyond the scope of this work. Given that the Hsig solution is not affected by these numerical artifacts and remains physically consistent across both spatial resolutions, we chose to partially minimize the discrepancies in the radiation stress gradients by isolating and eliminating the numerical artifacts limited to the regions along the domain boundaries through the application of a boundary truncation. Thus, the outermost boundary strip is removed from the evaluation domain as shown in Figure~\ref{fig:RefindementGrid}. Specifically, a buffer zone containing $1,957$ grid points in total was excluded from the evaluation domain. This truncated perimeter strip comprises 1 grid line (12.5 m) along the offshore boundary, 4 grid lines (50 m) along the nearshore boundary, 5 grid lines (250 m) along the top boundary, and 6 grid lines (300 m) along the bottom boundary. By removing these boundary region artifacts and evaluating model performance exclusively on the common grid points within the truncated interior domain, the discretization error between the coarse-mesh and fine-mesh SWAN baselines is effectively isolated from edge noise. When evaluated on the common grid points in the truncated interior domain, the mean $RLE^{(s)}$ between the coarse-mesh and fine-mesh SWAN results reduces to $0.61\%$ for Hsig, $4.25\%$ for x-direction, and 10.17\% for y-direction radiation stress gradients, as shown in Table~\ref{tab:ModelPerformance} (4th column). 

Using this truncated fine-mesh SWAN solution as the ground truth, the DeepONet surrogate is evaluated on these truncated fine-mesh coordinates. As detailed in Table~\ref{tab:ModelPerformance} (last column), the DeepONet prediction for Hsig on the completely unseen fine-mesh coordinates show a high degree of accuracy with a mean $RLE^{(s)}$ of 1.25\%, and the worst $RLE^{(s)}$ being 3.17\%, both of which are within the range of discrepancies between the coarse and fine-mesh SWAN solutions. Moreover, even the DeepONet predictions for the x- and y-direction radiation stress gradients remain reasonably bounded around the coarse and fine-mesh SWAN discrepancies for each variable, with a mean $RLE^{(s)}$ of $9.13\%$ and $13.95\%$, respectively. 

\begin{table*}[h]
\centering
\begin{tabular}{ccccc}
\textbf{Variable} & \textbf{Metrics} & 
\textbf{Full Mesh} & \textbf{Truncated Mesh} &
\textbf{DeepONet}  \\\midrule
\textbf{Hsig}
& Mean $RLE^{(s)}$
& 0.93\% & 0.61\% & 1.25\%  \\
& Median $RLE^{(s)}$
& 0.79\% & 0.48\% & 1.09\%   \\
& Worst $RLE^{(s)}$
& 2.27\% & 1.40\% & 3.17\% \\\midrule
\textbf{x-forces}
& Mean $RLE^{(s)}$
& 6.50\% & 4.25\% & 9.13\% \\
& Median $RLE^{(s)}$
& 6.00\% & 4.11\% & 8.67\%  \\
& Worst $RLE^{(s)}$
& 10.47\% & 7.01\% & 14.85\%  \\\midrule
\textbf{y-forces}
& Mean $RLE^{(s)}$
& 18.11\% & 10.17\% & 13.95\%   \\
& Median $RLE^{(s)}$
& 16.98\% & 9.04\% & 12.76\%   \\
& Worst $RLE^{(s)}$
& 29.47\% & 17.18\% & 21.04\%   \\
\end{tabular}
\caption{Cross-grid verification error metrics across 576 DUCK Test scenarios. The coarse-mesh SWAN results are compared with the fine-mesh SWAN results on the common grid points across (a) the entire fine mesh (column 3) and (b) the boundary-truncated fine-mesh (column 4). The last column shows the error metrics between the fine-mesh SWAN results and the DeepONet inference on the common grid points in the boundary-truncated fine-mesh.}
\label{tab:ModelPerformance}
\end{table*}

\begin{figure}[h!]
\centering
\includegraphics[width=0.9\textwidth]{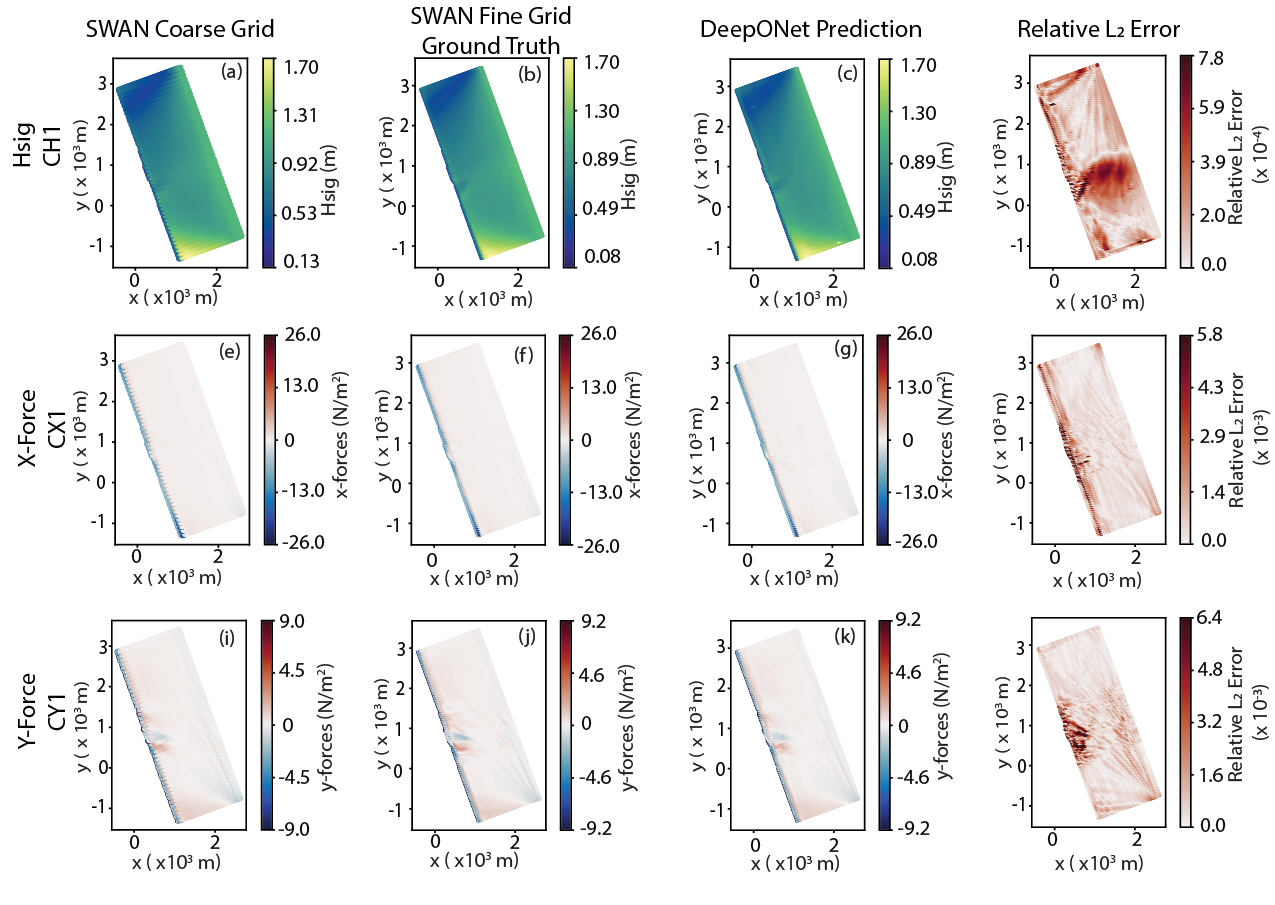}
\caption{Comparison between SWAN and DeepONet for the worst-case scenarios in the DUCK example with different mesh resolutions. For visualization purposes, the color bar of $RLE$ is clipped at the 99th percentile to reduce the influence of extreme values and better highlight the overall spatial error patterns.}
\label{fig:DUCK_CrossGrid}
\end{figure}

Table~\ref{tab:duck_crossGrid_worst_cases} shows the worst-performing test scenarios for each model (designated as scenarios CH1, CX1, and CY1). Figure~\ref{fig:DUCK_CrossGrid} presents the corresponding spatial distribution of Hsig, x-, and y-forces for these worst-case scenarios. Although these cases exhibit the largest $RLE^{(s)}$ among all test scenarios, a direct comparison between the fine-grid SWAN solutions (column 2) and the DeepONet predictions (column 3), for Hsig (row 1), x-forces (row 2), and y-forces (row 3), reveals that the macro-scale spatial patterns are in agreement. The location of major gradients and regions of high and low intensity are well reproduced by the DeepONet model. These results indicate that the model successfully learns the underlying continuous operator map, rather than merely memorizing local grid-dependent correlations. 

\begin{table*}[h]
\centering
\begin{tabular}{cccccc}
\textbf{Variable} & \textbf{Scenario} & 
\textbf{Wind} & \textbf{Wind Dir} &
\textbf{Wave H.} & \textbf{Wave Dir}  \\\midrule
\textbf{Hsig}
& CH1
& 7 & 315 & 1.6 & -20  \\
\midrule
\textbf{x-forces}
& CX1
& 8 & 0 & 1.6 & 160 \\
\midrule
\textbf{y-forces}
& CY1
& 7 & 315 & 2.6 & 160  \\
\end{tabular}
\caption{Scenario conditions corresponding to the worst-performing test scenarios for each variable in the DUCK cross-grid verification. Wind speed is given in m/s, wave height in m, and directional quantities in degrees. }

\label{tab:duck_crossGrid_worst_cases}
\end{table*}

Overall, the cross-grid verification demonstrates that the proposed DeepONet successfully generalizes beyond the spatial discretization used during training. Although the surrogate exhibits slightly larger errors than the numerical discrepancies introduced by mesh refinement in SWAN, the prediction errors remain within the same order of magnitude. This indicates that the majority of the observed differences can be largely attributed to the numerical solution's inherent sensitivity rather than to limitations of the learned operator. More importantly, the model maintains its predictive performance on previously unseen coordinate locations without retraining or interpolation, providing strong empirical evidence of the discretization-invariance of the DeepONet framework. These results suggest that the surrogate can be deployed on alternative mesh resolutions and spatial sampling patterns while preserving physically consistent predictions.

\section{Discussion}
\label{sec4}

This study evaluates the ability of DeepONets to approximate the nonlinear operator that maps wind and boundary wave conditions to steady-state bulk wave parameters such as the significant wave height and the radiation-stress gradients in coastal wave models. Across progressively more complex numerical examples: 1-D uniform slope and the realistic DUCK bathymetry, the surrogate demonstrates strong predictive skill and robust generalization. Several key insights emerge from these results.

\subsection{Hardware Profile and Computational Efficiency}
To evaluate the practical computational acceleration achieved by the DeepONet framework, an execution benchmark is conducted using 576 test scenarios from the DUCK example, comparing the standard numerical SWAN model with the DeepONet surrogate.

The DeepONet models were trained on the Texas Advanced Computing Center (TACC) Vista supercomputing cluster, utilizing a single NVIDIA GH200 Grace Hopper Superchip node. The training times required for each model are detailed in Table~\ref{tab:training_time}. 

\begin{table}[ht]
\centering
% \resizebox{\textwidth}{!}{
\begin{tabular}{c c c c}
    \toprule
    &  Model & 1-D Example & DUCK Example \\
    \midrule
    & Hsig   & $26~m~41~s$ & $12~h~15~m~33~s$ \\
    & x-forces   & $27~m~06~s$ &$14~h~04~m~34~s$ \\
    & y-forces   & -  &$14~h~02~m~58~s$ \\
    \bottomrule
\end{tabular} 
% }
\caption{Training times for the Hsig, x-forces, and y-forces DeepONet models of the 2-D DUCK example on the TACC Vista cluster (NVIDIA GH200)}\label{tab:training_time}%
\end{table}

\begin{figure}[h!]  
\centering
\includegraphics[width=0.9\textwidth]{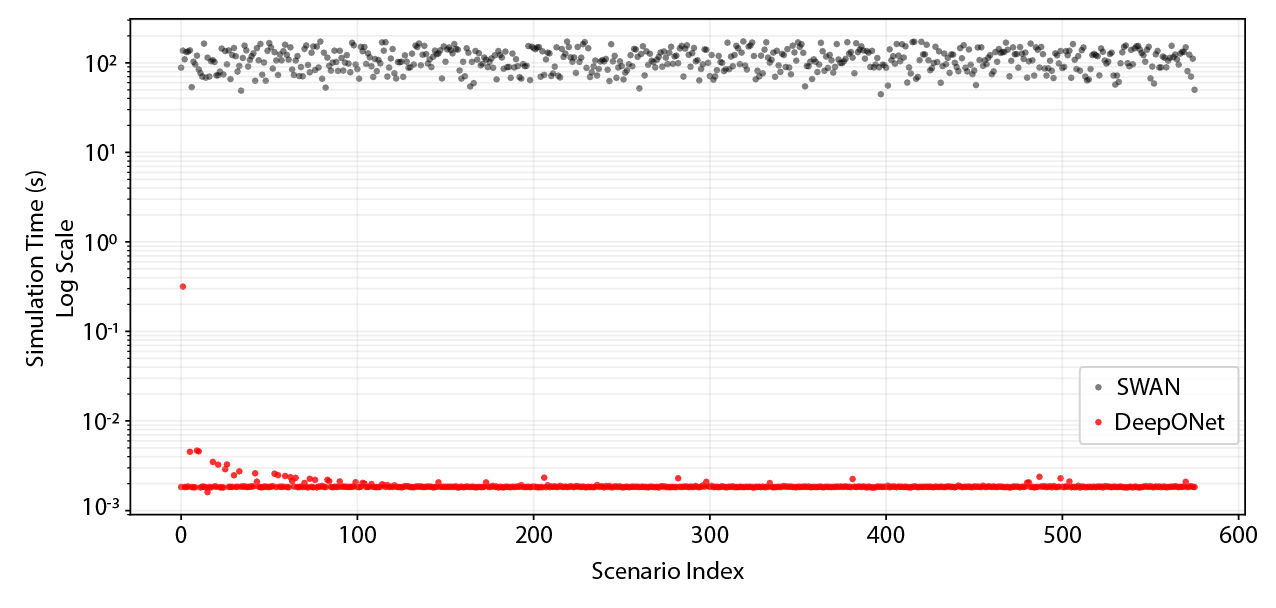}  
\caption{Scenarios-by-scenario computational runtime comparison between the SWAN model and DeepONet across all 576 test cases.}
\label{fig:ExecutionTime} 
\end{figure}

Model inference and numerical performance benchmark studies were conducted in a lab server environment. This local architecture features a dual-socket Intel Xeon Silver 4510 processor and an Nvidia L40S GPU. Figure~\ref{fig:ExecutionTime} illustrates the scenario-by-scenario execution time for both models across all test cases. The DeepONet demonstrates massive efficiency gain: while an individual SWAN run requires an execution runtime on the order of $O(10^2)$ seconds, the DeepONet inference completes within an order of $O(10^{-2})$ seconds. This amounts to a net computational speedup of approximately four orders of magnitude ($O(10^4)$).
\subsection{Model Performance Under Increasing Physical Complexity}

The DeepONet surrogate achieves high accuracy in both examples, with errors remaining small relative to the magnitude of the target variables. The surrogate's ability to reproduce spatial patterns in both Hsig and radiation stress gradients suggests that the network captures key physical processes such as shoaling and depth-induced breaking. Notably, even in the DUCK case, where bathymetry is irregular and nonlinear interactions dominate, the DeepONet surrogate models maintain consistent performance, highlighting DeepONet's capacity to approximate operators defined on complex domains. Beyond global performance evaluations, understanding where and why the model struggles provides deeper insight into its physical consistency.

\subsection{Spatial Localization of Errors and Physical Interpretability}

In every experiment, the largest absolute and mean errors occur within physically energetic regions: the surf zone where waves experience rapid shoaling and breaking and near offshore boundaries where spectral conditions are imposed. These regions correspond to locations where the governing equations exhibit heightened sensitivity and strong spatial gradients. The bottom row of Figure~\ref{fig:duck_mae} further shows that the points with the highest absolute error in each test scenario consistently cluster near the shoreline, demonstrating that error amplification is not random but tied to physical processes. Importantly, these local error concentrations do not undermine the overall accuracy of the predictions, as they represent a small fraction of the domain and occur where SWAN solutions themselves undergo abrupt spatial transitions.

\begin{figure}
    \centering
    \includegraphics[width=0.9\linewidth]{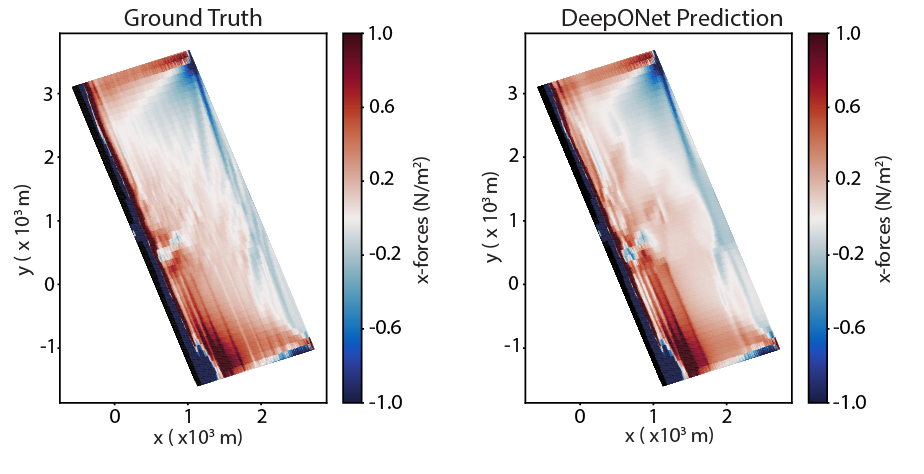}
    \caption{Spatial distribution of predicted x-forces compared with the benchmark SWAN under 7 m/s at $90^{\circ}$ and boundary wave height 1.6 m at $20^{\circ}$. The color bar is clipped between -1 and 1 $N/m^2$ to highlight spatial distribution patterns, illustrating DeepONet's spatial smoothing relative to the localized fluctuations in SWAN.}
    \label{fig:SpatialContinuous}
\end{figure}

A key observation across all numerical examples is the tendency of the DeepONet surrogate to smooth out high-frequency spatial oscillation in the SWAN output, as illustrated in Figure~\ref{fig:SpatialContinuous}. In the Duck example, the numerical solver produces high-frequency spatial fluctuations in the radiation-stress gradients, likely resulting from small-scale bathymetric irregularities and numerical discretization errors. While these "jagged" patterns appear in the ground truth, the DeepONet model produces a more continuous and regularized spatial distribution.

This behavior is likely due to spectral bias, in which neural networks prioritize learning low-frequency global patterns over high-frequency local variations. In regions of sharp physical transitions, such as the sandbars in Duck, this smoothing effect results in localized increases in relative $L_2$ error, as the surrogate fails to capture the exact magnitude of numerical "spikes." However, it is important to consider that many of these high-frequency fluctuations in SWAN may be numerical artifacts rather than physical signals. By providing a smoother, physically consistent gradient, the DeepONet surrogate may actually offer a more stable forcing term for coupled hydrodynamic circulation models, which are notoriously sensitive to the high-frequency numerical noise often present in traditional wave model outputs.

Despite localized discrepancies in the surf zone, the surrogate's ability to capture the global structure of radiation stress gradients enables seamless integration into broader, multi-scale modeling frameworks.

\subsection{Discretization Sensitivity}
The cross-grid verification study reveals the contrast in sensitivity between the Hsig field and radiation stress gradient fields. The Hsig simulation showed a low cross-grid discrepancy of $1.25\%$ whereas the discrepancies for the x- and y-direction radiation stress gradient fields were $9.13\%$ and $13.95\%$, respectively. This difference reflects the underlying physical properties of the variables themselves. 

Hsig is derived by integrating the continuous wave energy density spectrum over frequency and direction. By accumulating energy across the entire spectrum at a given spatial location, the integration process tends to smooth localized grid anomalies and high-frequency numerical noise introduced by transitioning to a refined mesh. In contrast, the radiation stress gradients are computed from spatial gradients of the radiation stress tensor and are, therefore, inherently more sensitive to grid resolution.
Furthermore, the greater discrepancies observed in the y-forces than in the x-forces may be linked to the spatial grid anisotropy of the F71 testbed. In nearshore wave modeling, cross-shore dynamics are often prioritized, resulting in a finer mesh resolution along the direction of the dominant wave-shoaling gradient. 

These results demonstrate the trained DeepONet surrogate's ability to perform inference even on a finer mesh than that used during training. The low discrepancy observed for Hsig suggests that the DeepONet learns the underlying continuous physical operator rather than simply memorizing fixed grid locations. However, for variables that depend on spatial derivatives, discrepancies are expected to be larger.

\subsection{Implications for Coupled Wave-Current Modeling}

The accurate prediction of radiation-stress gradients is particularly important because these quantities directly force circulation models such as ADCIRC. The surrogate model demonstrates the ability to approximate these gradients with high fidelity at an extremely low computational cost, providing a nearly four-order-of-magnitude speedup over traditional SWAN simulations. This substantial reduction in computational expense highlights the potential of the proposed approach to enable efficient wave–current coupling in large-scale or ensemble-based coastal simulations.

However, a primary challenge in integrating this surrogate into coupled wave–current frameworks lies in the high-frequency features present in the force gradients. While the surrogate successfully captures the underlying physical processes of wave-induced momentum flux, the "smoothing" effect of the DeepONet model may lead to discrepancies in regions where values are near zero or where the ground-truth solutions exhibit jagged, high-frequency patterns. Understanding the implications of this regularization is essential as it may influence the precision of the resulting current fields in traditional coupled modeling environments.

\subsection{Limitations and Future Work}
A fundamental limitation of data-driven surrogate models is that their predictions remain bounded by the underlying numerical solver and training data. When trained on coarse-mesh SWAN simulations, DeepONet does not learn physical processes that were not accounted for in the SWAN simulation or fine-scale bathymetric features that were absent in the coarse-grid input. Consequently, the surrogate model should be treated as an efficient operator approximation of the numerical solver used to generate the training data rather than the true governing physics equations. 

Furthermore, the input data resolution is also a limitation. A finer mesh would theoretically allow DeepONet to resolve more subtle physical features. However, the data generation may introduce a high computational cost. For instance, simulating the $576$ test scenarios of the DUCK example requires $17.72$ hours using the coarse mesh and $36.96$ hours using the fine mesh. Therefore, the surrogate model’s fine-scale predictive skill remains bounded by the computational cost of generating the requisite quality of numerical training data. 

The DeepONet surrogate models predict bulk wave properties but not the full action density spectrum. Extending this approach to spectral outputs would increase the dimensionality of the problem and may require more sophisticated architectures. Nevertheless, this work establishes a robust and promising framework for neural operator applications in coastal wave modeling, with implications summarized in Section~\ref{sec5}.

\section{Conclusion}
\label{sec5}

This work demonstrates that DeepONet can serve as an efficient and accurate surrogate for steady-state simulations of the SWAN spectral wave model in different coastal environments. By evaluating the surrogate using both a 1-D and a realistic 2-D DUCK example, we show that the DeepONet successfully learns the nonlinear operator that maps wind and boundary wave conditions to the significant wave height and the gradients of the radiation stress, three bulk wave parameters that are essential for capturing wave-current interactions in coastal systems. 

These results also highlight the potential of such neural operator models in accelerating process-based coupled wave-current modeling frameworks such as SWAN+ADCIRC. Unlike standard MLPs that are constrained to a fixed grid layout, a DeepONet model trained on only $5\%$ of the spatial grid points is shown to construct a continuous solution on the full domain, and even outperform a MLP of similar capacity that has been trained of the full grid solution, thus demonstrating strong cross-grid generalization capabilities. Additional, refined cross-grid verification results demonstrate that the surrogate can be evaluated on meshes different from those used during training while preserving predictive skill. Although larger discrepancies were observed for derivative-based x- and y-force predictions, these differences were shown to be consistent with the inherent sensitivity of the SWAN solver to mesh resolution rather than limitations of the neural operator model. This behavior suggests that the DeepONet learns the underlying continuous physical operator instead of merely memorizing grid-dependent solutions, an essential requirement for future deployment in coupled coastal forecasting systems.

At the same time, the present model focuses on bulk wave properties and assumes uniform wind forcing. Extending the surrogate model to predict full action density spectra, to incorporate time-varying or spatially heterogeneous wind fields, and to generalize across diverse coastal morphologies represent important directions for future research. 

Overall, this study provides a foundation for fast, data-driven surrogate modeling of coastal wave dynamics and demonstrates the viability of neural operator methods for accelerating or augmenting computationally expensive components of traditional coastal ocean forecasting systems. 

\section*{Data Statement}

The data generated using SWAN and trained DeepONet models will be made available via DesignSafe. The source code for data generation, model training and analysis will be made available at~\url{https://github.com/ShukaiC/NeuralOperator-CoastalWaves} upon publication of the article. 

\section*{Acknowledgements}

The authors would like to acknowledge the support from the US Department of Energy (DE-SC0022211), the U.S Army Engineer Research and Development Center (ERDC) through a Broad Agency Announcement (BAA W912HZ-23-2-0013) award, and the 2023 Laboratory-University Collaboration Initiative (LUCI) program, through an award made by the Office of the Under Secretary of War for Research and Engineering (OUSW(R\&E)), Science and Technology (S\&T) Foundations. The authors also gratefully acknowledge the computational resources provided by the Texas Advanced Computing Center (TACC) at The University of Texas at Austin, specifically the Frontera supercomputer under allocation DMS23001.

% \section{Example Appendix Section}


%%       \cite{<label>} ==> [1]

%%
% Example citation, See \cite{lamport94}.

\bibliographystyle{elsarticle-num} 
\bibliography{references}





% %% For numbered reference style

%   \textit{\LaTeX: a document preparation system},
%   Addison Wesley, Massachusetts,

% \end{thebibliography}
\end{document}